\documentstyle[aps,epsf,prb
  ]{revtex}
\begin{document}
\twocolumn[\hsize\textwidth\columnwidth\hsize\csname%
  @twocolumnfalse\endcsname
  
  \title{\strut{\vadjust{\vskip-0.5in\strut\hfill{\normalsize
          IASSNS-HEP-99/46\vskip0.5in}}} %
    Topological doping and the stability of stripe phases.}

\author{Leonid P.~Pryadko$^*$, Steven A.~Kivelson$^\dagger$,
  V.~J.~Emery$^\ddagger$, Yaroslaw B.~Bazaliy,$^{\mathchar "278}$ 
  and Eugene A.~Demler$^{\mathchar "27B}$}
\address{$^*$Institute for Advanced Study, Princeton, NJ 08540}
\address{$^\dagger$Dept. of Physics \& Astronomy, University of California,
  Los Angeles, CA 90095}
\address{$^\ddagger$ Dept. of Physics, Brookhaven National Laboratory,
  Upton, NY 11973-5000}  
\address{$^{\mathchar "278}$Dept. of Physics, Stanford University,
                 Stanford, CA 94305}
\address{$^{\mathchar "27B}$Institute for Theoretical Physics,
  University of California, Santa Barbara, CA 93106-4030}

\date{May 11, 1999}

\maketitle

\begin{abstract}
  We analyze the properties of a general Ginzburg-Landau free energy
  with competing order parameters, long-range interactions, and global
  constraints ({\it e.g.}, a fixed value of a total ``charge'') to
  address the physics of stripe phases in underdoped high-$T_c$ and
  related materials.  For a local free energy limited to quadratic
  terms of the gradient expansion, only uniform or phase-separated
  configurations are thermodynamically stable.  ``Stripe'' or other
  non-uniform phases can be stabilized by long-range forces, but can
  only have non-topological (in-phase) domain walls where the
  components of the antiferromagnetic order parameter never change
  sign, and the periods of charge and spin density waves coincide. The
  {\em antiphase\/} domain walls observed experimentally require
  physics on an intermediate lengthscale, and they are absent from a
  model that involves only long-distance physics. Dense stripe phases
  can be stable even in the absence of long-range forces, but domain
  walls always attract at large distances, {\em i.e.}, there is a
  ubiquitous tendency to phase separation at small doping.  The
  implications for the phase diagram of underdoped cuprates are
  discussed.
\end{abstract}
]
\narrowtext
\section*{Introduction}
One of the fundamental issues in the theory of highly correlated
solids is the nature of the ground-state phases produced when a small
concentration, $x$, of ``doped holes'' is introduced into a Mott
insulator, particularly an antiferromagnet.  It is now
established\cite{PKH-98,Hellberg-97,Hellberg-98,Kivelson-LosAlamos} that, at
small enough $x$ and in the absence of long-range Coulomb
interactions, a doped antiferromagnet generally phase separates into a
hole-rich and a hole-free phase, {\it i.e.\/}\ the antiferromagnetic
state is destroyed via a first order phase transition.  In the
presence of weak, long-range Coulomb interactions, that frustrate
this local tendency to phase separation, the two-phase region is
replaced by states which are inhomogeneous on intermediate length
scales,\cite{Kivelson-LosAlamos,Low-94,Andelman}
and especially ``stripe phases,''  which have now been
observed in a wide variety of oxide
materials.\cite{Tranquada-98,Hwang-97,Mori-98,Birgeneau-Oxigen,%
  Yamada-Elastic} 
In various quasi two-dimensional cuprate high temperature
superconductors and the isostructural nickelates the stripes are
observed\cite{Tranquada-97} to be ``topological,'' in the sense
that the charge is concentrated along one-dimensional ``rivers'' which
are at the same time antiphase domain walls in the antiferromagnetic
order.  In the nearly cubic manganate colossal magnetoresistance
materials,\cite{Hwang-97,Mori-98} the ``stripes'' are two dimensional
sheets of charge which are non-topological. (In some sense, each sheet
can be thought of as a dimer of topological
stripes.\cite{Mori-98,endn:one})

Here we study the properties of a general
Ginzburg-Landau free energy with competing order parameters,
long-range interactions, and global constraints [{\it e.g.}, a fixed
  value of a total ``charge'', as defined in
  Eq.~(\ref{eq:charge-constraint})] to address the physics of
inhomogeneous (``stripe'') phases.  Specifically, a
stripe phase is a unidirectional density wave which, in
the case of a doped antiferromagnet, consists of a coupled spin-density
wave (SDW) and charge-density wave (CDW).  At very dilute doping, a
stripe phase consists of an ordered array of far-separated
self-localized structures, or individual stripes.  At moderate doping
levels, where the spacing between stripes is comparable to their
width, the structures are best described as nearly harmonic density
waves.

Zachar and two of us\cite{Zachar-97} have considered the density wave limit of
a Landau theory of coupled CDW and SDW
order, {\it each with a fixed wave vector $\vec{q}$},
near a transition to a disordered state, that occurs as the
temperature or doping  
are varied. The existence of a cubic term in the Landau
free energy coupling these two order parameters 
drives the period of the SDW to be twice that of the CDW, and
the absence of any net AF ordering is equivalent to the statement that
the stripes are topological.  By contrast, as shown in
Appendix~\ref{density-func}, the same sort of term in the Landau
theory of the transition between a homogeneous ordered antiferromagnetic phase
and a stripe ordered phase produces a
state in which the antiferromagnetic magnetization {\em does not\/}
change its sign between the domains, {\it i.e.\/}\ the stripes are
non-topological. 

To elucidate the circumstances in which arrays of stripes can be
thermodynamically stable, and what determines their character ({\it
  i.e.\/}\ topological {\em vs.\/}\ non-topological, collinear {\em
  vs}.\ spiral) we shall
concentrate on the dilute limit, where the spacing between stripes is
large, and the stripes are highly anharmonic structures.
Specifically, we study the extremal states of a general
Ginzburg-Landau free energy functional for coupled order parameters
as a function of the average charge density.

Whenever the order parameter profiles are {\em slowly varying
everywhere}, so that only the lowest order (quadratic) terms in the
gradient expansion of the free energy are necessary
[Eqs.~(\ref{eq:most-general-cont}) and
(\ref{eq:long-distance-interaction})], we show that:

{\noindent 1)} In the absence of long-range interactions, only
spatially uniform and phase separated (two-phase coexistence) states
are globally stable.

{\noindent 2)} ``Stripe'' or other non-uniform phases can be
stabilized by long-range forces, but they are non topological in the
sense that any component $u_{i}$ of the order parameter has a uniform
sign so long as the free energy density is an even function of
$u_{i}$.  [We indicate all point symmetry groups which satisfy this
condition for a magnetic (pseudovector) order parameter.]

{\noindent 3)} Whenever there is a global rotational symmetry of the
order parameter, any localized configuration which interpolates
between two distinct asymptotic ground states ({\it e.g.\/}\ an
antiphase domain wall) is locally unstable to untwisting.

The possibilities become richer in cases in which higher order
derivative terms in the Ginzburg-Landau free energy or lattice effects
determine an additional length scale---the core size of a localized
defect.\cite{Zachar-97} When there is no frustration, topological
stripes are still forbidden in the ground state.  However,
frustration, such as competing first and second neighbor interactions
in a lattice model, or opposite sign terms in the gradient expansion
of the Ginzburg-Landau model ({\it i.e.\/}\ below a Lifshitz point),
can stabilize topological collinear domain walls.  
In the context of
doped antiferromagnets, this kind of frustration can arise as a result of the
competition between the tendency of the Coulomb interaction to
localize the charges, and the tendency of electrons to quantum
delocalize.  However, even in this case, the asymptotic interaction
between defects is still attractive at large distances, so long-range
forces are necessary to suppress phase separation in the dilute limit.

In other words, topological stripes are a consequence of physics on an
intermediate lengthscale, and they do not appear in a theory that
considers only 
long-distance or low-energy physics.

The plan of this paper is as follows.  In Sec.~\ref{sec:background} we
review some of the theoretical and experimental background. 
Specifically, we discuss some of the early theoretical work
predicting stripe phases, the theoretical controversies concerning the
range of phase separation in microscopic models, such as the $t-J$
model, and some of the experimental facts concerning stripe phases in
doped antiferromagnetic insulators.

In Sec.~\ref{sec:phase-separ} we perform a scaling analysis of
possible non-uniform configurations which minimize a generalized
Ginzburg-Landau 
functional, establish the analog of the virial theorem which relates
the long-distance Coulomb interaction to  the gradient energies of the
system, and 
derive the universal asymptotic form of the large distance
interactions between domain walls or other defects.

In Sec.~\ref{sec:no-twist} we analyze the local and global stability
of non-uniform ground state configurations.  For systems with a global
rotational symmetry of the order parameter, we show that the antiphase
domain walls are locally unstable to ``untwisting'', even in the
presence of long-range forces.  If the rotational symmetry is broken
these domain walls can be locally stable, but they are not necessarily
allowed in any ground state configuration.  We establish a
corresponding {\em sufficiency criterion for global instability\/} for
such antiphase domain walls, and identify the corresponding point
symmetry groups of the underlying lattice.

In Sec.~\ref{sec:antiphase}, we show that antiphase domain walls can
be stable even in the ground state, if the free energy functional
includes higher derivative terms or is defined on the lattice.  We
discuss a sufficiency criterion for local stability of the solutions,
and illustrate the effect of stabilization of antiphase domain walls
in particular examples.  
We also show that, for systems with
short-range interactions and mixed AF and charged order parameters,
the domain walls always attract at large distances, which indicates a
tendency to phase separation at small doping.
If long-range Coulomb interactions are included as well, 
inhomogeneous phases are stabilized.
Depending on details, 
either wide stripes are produced via
Coulomb-frustrated phase separation\cite{Emery-93,Low-94}, or 
certain dense 
stripe phases are stabilized, in agreement with the
arguments of Hellberg and Manousakis\cite{Hellberg-99,Kivelson-LosAlamos}.

We conclude that although (avoided) phase separation is ubiquitous,
especially at small doping, antiphase domain
walls are not universal in the ground state, even in the presence of
long-range forces.
Certain types of short-distance physics are required to stabilize
antiphase domain walls. Therefore, effective long-distance models are
not, in general, sufficient for successful description of the stripe
morphology in the
cuprates and nickelates.

\section{background}
\label{sec:background}

The undoped parent compounds of the high-$T_c$ materials have one
electron per unit lattice cell, and, if it were not for the
electron-electron interactions, one would expect them to be metallic.
Instead, strong Coulomb repulsion renders the system a Mott insulator
and results in an antiferromagnetic (AF) ground state with a doubled
unit cell.  Unlike usual band insulators, such {\em correlated
  insulators\/} do not conduct even when weakly doped.  The
short-distance physics of the doped system, dominated by strong
electron-electron repulsion, is believed to be captured in the
large-$U$ Hubbard model, the $t$-$J$ model\cite{Anderson-87}, or
related models.\cite{Emery-87}

Unfortunately, to this time, none of these models has been solved in
anything resembling a physical regime of parameters.  One well
established aspect is the tendency of these models to phase
separation\cite{Visscher-74,Emery-90,Marder-90,%
  PKH-98,Hellberg-97,Hellberg-98,Kivelson-LosAlamos} in a substantial
range of 
parameters.  In the presence of the long-range Coulomb repulsion phase
separation is, of course, impossible, unless the dopants are mobile.  
Instead, the system forms a charge-inhomogeneous state, in which
hole-rich regions
exist in an antiferromagnetic background.\cite{Kivelson-LosAlamos,Low-94}
Within this picture, it is natural to interpret the stripe phases
observed in various doped antiferromagnets as being a consequence of
Coulomb-frustrated electronic phase separation (sometimes called
micro-phase separation.\cite{Mori-98}) Such stripe phases can
be either metallic or insulating, depending on the character of the
hole-rich phase.\cite{Kivelson-LosAlamos,Kivelson-topo-96,Nayak-97,%
  White-98A,White-98B,Kivelson-98}
However, the precise range of parameters in which phase separation
occurs in systems with short-range interactions, and even the physical
reasons for the stability of antiphase domain walls in systems with
Heisenberg symmetry, have not been fully elucidated. Moreover, phase
separation, especially at small doping, is notoriously hard to see
numerically; even for the most studied $t-J$ model, some numerical
studies have been interpreted as indicative
of\cite{Emery-90,Hellberg-97,Hellberg-98,Calandra-98} the universality
of phase 
separation in the limit of small doping, while others purport to 
indicate the existence of a critical ratio of $J/t$ below which phase
separation does not occur.\cite{Rice-94,White-98A,White-98B}

For the case of doped AFs with unbroken spin-ro\-ta\-tional invariance
this controversy was resolved by Pryadko, Kivelson and
Hone\cite{PKH-98}.  It was shown that spin-wave exchange always causes
an attraction between localized holes or hole clusters, similar to the
well-known Casimir effect\cite{Casimir-48}.  At large distances this
attraction falls off as a power law, and therefore it is always
stronger then the exponentially-decreasing forces present in the
system with short-range interactions.  This proves that any phase with
static charge order is thermodynamically unstable at small enough
doping.  However, the absolute magnitude of this attractive force is
very small, and even a relatively weak easy-axis anisotropy (allowed
by the symmetry in planar materials) can provide a spin-wave gap
sufficient to suppress this effect.

Static incommensurate magnetic and charge order in the cuprate high
temperature superconductors was first discovered \cite{Tranquada-95}
in La$_{1.6-x}$Nd$_{0.4}$Sr$_x$Cu\,O$_{4+\delta}$.  
Recently,
X-ray\cite{vonZimmermann-97,Vigliante-97,vonZimmermann-98} diffraction
measurements have confirmed the existence of charge order. 
Moreover, in this
material, static stripe order coexists\cite{Tranquada-97B,Ostenson-97}
with superconductivity, albeit with suppressed $T_{c}$.  Additional
indirect information about the frequency range of magnetic
correlations was provided by local probes, such as $\mu$-SR
\cite{Luke-97,Wagener-97B,Nachumi-98}.
In this material a structural phase transition to a low temperature
tetragonal (LTT) phase substantially stabilizes the stripe order,
making it particularly easy to detect, but, at the same time,
suppresses the superconducting transition temperatures.  Indeed, in
closely related materials ({\it e.g.\/}\ %
La$_{1.4-x}$Nd$_{0.6}$Sr$_x$Cu\,O$_{4}$), static stripe order is 
observed, but no evidence of
superconductivity has been found.\cite{Buchner-93,Buchner-94}
However, more recently, static stripe order has been
detected\cite{Yamada-Elastic} in the more widely studied high
temperature superconductors La$_{2-x}$Sr$_x$Cu\,O$_{4}$ with $0.5 < x
< 0.13$ and\cite{Birgeneau-Oxigen} ``stage-IV''
La$_{2}$Cu\,O$_{4+\delta}$, in which the transition temperature
$T_{c}=42K$ is not suppressed. 

Moreover, evidence has mounted that in a still broader class of high
temperature superconductors (perhaps even all high temperature
superconductors) stripe order is nearly condensed in the sense that
there are substantial stripe-like correlations which persist at low
temperatures over long intervals of space and time. Slow dynamically
fluctuating incommensurate magnetic correlations were observed some
time ago\cite{Cheong-91} by inelastic neutron scattering
in La$_{2-x}$Sr$_x$Cu\,O$_{4}$. 
That these incommensurate structures are simply fluctuating stripes is
now clear from a comparison\cite{Tranquada-95,Tranquada-98} of the
fluctuations in
this material and its ordered cousin,
La$_{1.6-x}$Nd$_{0.4}$Sr$_x$Cu\,O$_{4}$.  Evidence supporting the
universality of incommensurate fluctuations in high-$T_c$ materials
has also been recently provided by neutron scattering
studies\cite{Mook-98} of spin fluctuations in YBa$_2$Cu$_3$O$_{7-x}$
and Bi$_2$Sr$_2$CaCu$_2$O$_8$, and indirect evidence of the same
structures in Bi$_2$Sr$_2$CaCu$_2$O$_8$ has been obtained from angle
resolved photoemission (ARPES).\cite{Shen-98} 
Indirect evidence that
static stripe structures may also be more common than previously
appreciated can be deduced from $\mu$-SR
measurements\cite{Niedermayer-98} and NQR measurements.\cite{Imai-99}
The existence of stripe phases was first established  
in the nickelates (La$_{2-x}$Sr$_x$Ni\,O$_{4+\delta}$)
by direct electron\cite{Chen-93} and
neutron\cite{Tranquada-95A,Tranquada-96B,Lee-97} scattering.
But the ubiquity of stripe phases in doped antiferromagnets
has become clear only in the last couple of years of intensive
experimental inquiry.  
Stripe order in the insulating, nearly cubic
manganates has been vividly visualized by electron diffraction
studies.\cite{Mori-98} Here the charge order is strongly coupled to a
lattice (Jahn-Teller) distortion, which makes the stripes more
classical and more strongly ordered; the stripes here are
non-topological in the sense that the CDW period is equal to the SDW
period. The real-space images constructed from the electron
diffraction results make it clear that each non-topological stripe can
be viewed as a pair of close-by topological stripes, or equivalently
that the topological stripe array has been dimerized.

In all cases in the cuprates and nickelates, where the information is
available, the measured positions of the incommensurate peaks indicate
that the period of spin modulation is twice that of the charge
modulation.  This, and other data\cite{Tranquada-98}, support the
model\cite{Zaanen-89,Zaanen-94} of charged holes concentrated on the
antiphase walls between neighboring antiferromagnetic domains.  The
effect of stabilization of such antiphase domain walls, or stripes, by
the addition of charged holes to a correlated insulator, was
named\cite{Kivelson-topo-96} {\em topological doping}.

But while the existence of stripe phases in doped antiferromagnets is
clearly established, and there is growing evidence that it is a
general phenomenon, there is less agreement on the origins of the
stripes and their implications.
The existence of stripe phases consisting of arrays of antiphase
domain walls in doped antiferromagnets was, in fact, predicted still
earlier than the work\cite{Emery-93} on Coulomb frustrated phase
separation on the 
basis of Hartree-Fock mean-field theory.\cite{Zaanen-89,Schulz-89} The
Hartree-Fock stripes always have a commensurate density of holes
corresponding to one hole per site along the length of the stripes,
and are always insulating; a gap equal to a substantial fraction of
the insulating gap opens at the transition to the Hartree Fock stripe
phase.  These are generalizations of similar calculations in one
dimension\cite{Heeger-88} to the higher dimensional case, and are
closely related to calculations\cite{KLevin-91} which sought to explain
the existence of strong incommensurate peaks in the magnetic
susceptibility in terms of Fermi surface nesting; 
the stripe phase in
Hartree-Fock theory is directly a consequence of that
nesting.\cite{Schulz-89} In detail, these approaches do not account 
for the behavior of the cuprates, in which the density of holes along a
stripe varies\cite{See-Yamada-98} continuously as a function of $x$, and  
the stripe phases are conducting or superconducting,
not insulating.  Moreover, the evidence from ARPES is that there are
no sharply defined quasiparticles in the normal state of the
cuprates.\cite{Fedorov-99} 
In the LSCO family of materials, in which the evidence
of stripe order and stripe fluctuations is strongest, there is simply
no vestige of a quasiparticle in the region of momentum space
where the nested Fermi surface is supposed to occur.\cite{Shen-98}
However, these mean-field Hartree-Fock calculations already reflected
the tendency\cite{Schrieffer-88} of the holes to be collectively
self trapped in regions of suppressed antiferromagnetism, a close
relative of phase separation.  Moreover, they correctly identify the
microscopic physics, the transverse kinetic energy of the holes, which
gives rise to the anti-phase character of the stripes.

The unreliability of the Hartree-Fock approxi\-ma\-tion for
determining the properties of domain walls in strongly-coupled systems
was also pointed out by Nayak and Wilczek\cite{Nayak-97}.  They
analyzed the energy per electron on a partially filled stripe, which,
ignoring the effect of antiferromagnetic surrounding, was approximated
as the sum of the energy of broken AF bonds and the kinetic energy of
one-dimensional electrons in the limit $U\to\infty$.  Even in the
absence of long-range interactions, the model does not develop a gap,
and the value of the optimal filling of the stripes was shown to vary
continuously with parameters.  Therefore, the stripes in this
approximation are conducting and not insulating as follows from the
Hartree-Fock analysis.

An alternative phenomenology of high-$T_c$ materials was suggested by
S.-C. Zhang\cite{Zhang-science}, who emphasized the competition
between the superconducting and AF order parameters.  In the
vicinity of a (hypothetical) $SO(5)$ symmetric point, where these two
order parameters form a five-dimensional vector of ``superspin'', the
effective free energy can be written in general Ginzburg-Landau form,
with relatively small symmetry breaking terms.  
Analysis\cite{Bazaliy-stripes} of non-uniform MF solutions in such a
model (assuming that the magnitude of the five-dimensional
``superspin'' remains constant) was recently performed by M.~Veillette
{\em et\ al}.  In the absence of the long-range Coulomb interaction,
and at small enough doping, a Maxwell construction was used to show
that the system phase separates into antiferromagnetic and
superconducting regions.  Turning on the long-distance Coulomb
interaction stabilizes a variety of non-uniform droplet and stripe
phases.  Surprisingly (at the time), the expected antiphase domain
walls were {\em not\/} discovered among the numerical solutions.  The
signs of both AF and SC order parameters were {\em always\/} uniform,
although their magnitude changed substantially.  It is apparent that
the absence of antiphase domain walls is an artifact of the 
model, but the specific reason for this feature was not elucidated.

\section{Mixed phase or phase separation?}
\label{sec:phase-separ}
\subsection{General scaling arguments.}
\label{sec:scaling}

The mean field approach typically works well if the important degrees
of freedom vary slowly in time and space.  In such cases one can write
an effective free energy in generalized Ginzburg-Landau form
\begin{equation}
  \label{eq:most-general-cont}
  {\cal F}_l=\int d^Dx\left\{ \sum_i\left[ \chi_i({\bf u})\,(\nabla
  u_i)^2\right] 
  +V({\bf u}) \right\},
\end{equation}
which retains only the leading (quadratic) terms in the expansion over
the gradients of the order parameters $u_i$.  Usually, such a form of
the free energy [with $\chi={\rm Const}$ and polynomial $V({\bf u})$]
is used in the vicinity of a second order phase transition, where the
selection of the important terms is dictated by their ``relevance'' in
the sense of an appropriate renormalization group flow.  Similarly, in
high-energy applications\cite{Dashen-74,Christ-75,Friedberg-76A}, only
renormalizable potentials are usually considered.  Here, we shall try
to make as general an analysis as possible, and only assume that the
positive susceptibilities $\chi_i({\bf u})$ and the
potential energy $V({\bf u})$, which is bounded from below, 
are smooth enough functions of their
arguments, so that a lowest energy configuration always exists.  Such a
generalization of the Ginzburg-Landau free energy 
functional is necessary because, as
we shall show, form~(\ref{eq:most-general-cont}) is {\em not\/}
sufficient for describing the stripe phases of interest, independently of
the specific form of the local potential $V$.

The first statement is that the ground state of the
model~(\ref{eq:most-general-cont}), possibly with one or more
constraints of the form
\begin{equation}
  \label{eq:charge-constraint}
  {\cal Q}=\int d^Dx\,\rho({\bf u}),
\end{equation}
is either uniform or phase-separated in the thermodynamic limit; {\em
  the energy of any mixed (non-uniform) phase can always be lowered in
  an infinite system}.  To prove this, let us imagine that it were not the
case and that some non-uniform configuration ${\bf u}={\bf u}^{(1)}({x})$
(which, generally, we can assume to be periodic) minimizes the free
energy density $f={\cal F}/\Omega$, and also, if necessary, satisfies
the constraint for the charge density $\bar\rho={\cal Q}/\Omega$.
Then the dilated fields, ${\bf u}^{(\lambda)}\equiv {\bf
  u}^{(1)}(\lambda\,x)$ satisfy the same constraints, while the
corresponding energy density
\begin{equation}
  \label{eq:short-range-scaling}
  f_\lambda=\lambda^2\,{\rm K}^{(1)}+\Pi^{(1)},
\end{equation}
written here in terms of the original ``kinetic'' and ``potential''
energy densities
$$
{\rm K}^{(1)}\equiv\int\! \sum_i\chi_i \left(\nabla
  u_i\right)^2{d^Dx\over\Omega},\quad \Pi^{(1)}\equiv \int 
V\left({\bf u}\right)\,{d^Dx\over\Omega},
$$
evaluated at the configurations ${\bf u}={\bf u}^{(1)}(x)$, 
can be reduced by decreasing the scale parameter $\lambda$, which
is equivalent to a uniform dilation of the original field
configuration.  This contradicts the original assumption, and we
conclude that no such coordinate-dependent configuration can minimize
the energy of the system.

It is important to emphasize that the statement proven above is only correct
in the thermodynamic limit.  For a periodic solution in a {\em
  finite\/} system the scaling parameter $\lambda$ can take only
discrete values, so that at least one period would fit the system
size.  Further energy density reduction is possible by doubling both
the system size and the total charge, and then performing an
additional rescaling.  
Such scaling also has a direct implication for possible numerical
studies of this and related models: because $\lambda^2\, {\rm
  K}^{(1)}\sim 1/L^2$, the finite-size correction to the free energy
and other parameters will be likely to fall off as a power of the
system size.

At first sight it appears that the existence of stable kinks for any
symmetric double-well potential contradicts this statement.  We must
point out, however, that only a single-kink solution is topologically
stable; in any configuration with periodic boundary conditions one has
an equal number of kinks and antikinks, and the energy can be lowered
by annihilating the pairs.  For periodic potentials, multi-kink
configurations may be topologically stable, as long as the total
number of kinks is fixed by the boundary conditions.  With free or
periodic boundary conditions, however, such extremal solutions never
represent the ground state of the system.

Similarly, one can create stable {\em non-topological\/}
sol\-i\-tons\cite{Friedberg-76A,Rajaraman-Weinberg-75,%
  Friedberg-76,Emery-76,Rajaraman-79,Shiff-82} by minimizing the
energy of the system with an imposed {\em finite charge\/},
as opposed to a {\em finite charge density}, constraint.  In this
case the amount of charge itself is used to introduce an additional
length scale which fixes the size of the soliton, and the question
about phase separation does not arise.  The solution of this apparent
paradox is that, if the thermodynamic limit is defined correctly, both
the energy~(\ref{eq:most-general-cont}) and the conserved
charge~(\ref{eq:charge-constraint}) will turn out to be infinite (or
zero), and they cannot be used to define a length scale.  Only in this
case the correct procedure is to minimize the finite {\em density\/}
of the system's free energy, at a given charge density.

Let us now consider how the scaling in
Eq.~(\ref{eq:short-range-scaling}) is modified in the presence of a
long-range interaction
\begin{equation}
  \label{eq:long-distance-interaction}
  {\cal F}_C=\!\int\! d^Dx\,d^Dx'\,{[\rho({\bf
    u}(x))-\bar\rho]\,[\rho({\bf u}({x'}))
  -\bar\rho]\over |x-x'|^{\gamma}},
\end{equation}
where $\gamma<D$ for convergence.
Obviously, in this case the total charge
constraint~(\ref{eq:charge-constraint}) can be dropped, because the
integration in Eq.~(\ref{eq:long-distance-interaction}) will diverge
in large systems if the screening is not perfect, no matter how weak
the interaction is.  Evaluating the free energy density along the
dilated field configuration $u^{(\lambda)}$ (which, of course, must
have the correct value of the average charge density, so that the
long-range part of the energy is finite) we obtain, instead of
Eq.~(\ref{eq:short-range-scaling}),
\begin{equation}
  \label{eq:long-range-scaling}
  f_\lambda=\lambda^2\,{\rm
    K}^{(1)}+\Pi^{(1)}+{\lambda^{-D+\gamma}}\,{\rm V}^{(1)},
\end{equation}
where ${\rm V}^{(1)}$ is the long-range
energy~(\ref{eq:long-distance-interaction}) per unit volume, evaluated
for the field 
configuration ${\bf u}^{(1)}$.  The
integral~(\ref{eq:long-distance-interaction}) converges if
$D-\gamma>0$, and the free energy density $f_\lambda$ has a minimum at
$\lambda=1$ if
\begin{equation}
  \label{eq:virial-long-range}
  2{\rm K}=(D-\gamma)\,{\rm V}.
\end{equation}
This expression is analogous to the virial
theorem\cite{Rajaraman-Book} for the considered class of models.  It
is the manifestation of the equilibrium between competing gradient
terms, which tend to dilate the system, and the long-range forces,
which tend to decrease the scale of charge variations.  As a result of
this competition, an additional length scale is introduced into the
problem, and periodic field configurations can be
stabilized.\cite{endn:two}

\subsection{Interaction of defects.}
\label{sec:interaction}

Despite its generality, the scaling technique, considered above, is
limited to continuous models.  Furthermore, it is not sensitive enough
for analyzing the stability of more general models,
where the existence of mixed phases may depend on actual parameters.
Indeed, if the shape of individual soliton- or instanton-like defects
 for a given model is fixed at some short scale, the
mixed phase can often be understood as a lattice of such relatively
weakly coupled defects.  The stability of such a phase will be defined
by sound-like displacement modes, which are likely to be much softer
then the uniform dilations we considered so far.  The relevant
elasticity modulus will obviously be defined by the interaction
between the constituent defects.

In this
section we discuss how the asymptotic form of interaction between widely
separated solitons can be found by a simple linear analysis,
even though the core structure of the solitons themselves is governed
by a complicated set of non-linear differential equations.
Qualitatively, this is so
because away from their cores solitons asymptotically approach
one of the uniform ``vacuum'' configurations, and the interaction
between two solitons, placed 
sufficiently far apart, can depend only on the form of this asymptotic
fall-off.  Indeed, the mutual interaction can be interpreted as a force
exerted on the core of either soliton in the presence of the
infinitesimal field created by the other; therefore, this interaction
cannot depend on the internal structure of either soliton as long as
the large-distance asymptotic form remains the same.

This implies that the interaction between individual solitons must be
totally determined by the region of overlapping tails.  In this region
the amplitude of the perturbation of the vacuum is small, and the
effective free energy can be linearized.  After this step, the
linearized problem
reduces to a static Schr\"odinger equation in an external potential,
and the interaction energy can be found by standard
methods\cite{LL-Quant}.

As an illustration\cite{endn:three}, consider a one-dimensional (D=1)
free energy of the form~(\ref{eq:most-general-cont}), with constant
susceptibilities $\chi_i=1/2$, and the potential $V({\bf u})\ge 0$
reaching global minima only at ${\bf u}_\pm=\pm {\bf m}$, $V(\pm {\bf
  m})=0$.  In the absence of any special symmetries, there exists only
one (up to translations) minimal-energy trajectory ${\bf u}^0(x)$
interpolating between these minima, ${\bf u}^0(\pm\infty)=\pm {\bf
  m}$.  With this trajectory, we can also construct the approximate
double-kink trajectories of the form
\begin{equation}
  {\bf u}(x)={\bf u}^0(x-x_1)+{\bf u}^0(x_2-x)-{\bf
    m},\label{eq:double-kink}
\end{equation}
and write the corresponding interaction energy as
\begin{eqnarray*}
  \delta{\cal F}&\equiv& {\cal F}[{\bf u}_1+{\bf u}_2-{\bf m}]-{\cal
    F}[{\bf
    u}_1]-{\cal F}[{\bf u}_2]  =\int_{-\infty}^\infty dx\\
  & & \times\bigl[ {\bf u}_1'\, {\bf u}_2' +V({\bf u}_1+{\bf u}_2-{\bf
    m})- V({\bf u}_1)-V({\bf u}_2)\bigr],
\end{eqnarray*}
where ${\bf u}_{1,2}={\bf u}^0(\pm x\mp x_{1,2})$, and the prime
denotes the spatial derivative.  Let us choose a point $x_0$ somewhere
between the positions of the kinks, $x_1\ll x_0\ll x_2$.  Then, in the
left domain, $x<x_0$, the field $\delta{\bf u}_1\equiv {\bf u}_2-{\bf
  m}$ is small and can be considered as a small perturbation, while in
the region $x>x_0$ the field $\delta{\bf u}_2\equiv{\bf u}_1-{\bf m}$
is small.  Keeping only the terms of linear order in each domain, we
obtain
\begin{eqnarray}
  \delta{\cal F}&=&\!\int_{-\infty}^{x_0} \!\!\!dx\biggl\{
  \left({\bf u}_1'\delta{\bf u}_1\right)'+\delta{\bf u}_1
  \left[{-{\bf u}_1''+{\partial\over \partial {\bf u}} V({\bf
        u}_1)}\right]\biggr\}
  \nonumber\\
  & & + \int_{x_0}^\infty\!\!\!dx\biggl\{1\leftrightarrow2\biggr\},
  \label{eq:double-kink-energ}
\end{eqnarray}
or just
\begin{equation}
  \label{eq:interaction-energy-simple}
  \delta{\cal F} =\left. {\bf u}_1'({\bf u}_2-{\bf m})-{\bf u}_2'({\bf
      u}_1-{\bf m})\right|_{x=x_0}, 
\end{equation}
where the bulk terms disappear to this order because each field, ${\bf
  u}_1$ and ${\bf u}_2$, obeys the Euler-Lagrange extremum equations
exactly.  Despite appearances, the interaction
energy~(\ref{eq:interaction-energy-simple}) is actually independent of
the choice of the point $x_0$, as long as it is located far enough from
the cores of the kinks, so that the linearized Euler-Lagrange
equations apply.

Equation~(\ref{eq:interaction-energy-simple}) relates the
long-distance interaction between the kink and the antikink with their
asymptotic form at large distances.  For multi-component order
parameters the
 asymptotic properties may vary.  However, in the
particular case of antisymmetric kinks, ${\bf u}_0(x)=-{\bf u}_0(-x)$, we
can choose the separation point $x_0=(x_2+x_1)/2$ exactly midway
between the kinks, and the interaction energy can be rewritten as
\begin{eqnarray*}
  \delta{\cal F}& =& 2\left. {\bf u}_0'({\bf u}_0-{\bf
      m})\right|_{x=L/2}\\
  & =&\left.{d\over dx} \left({\bf u}_0-{\bf m}\right)^2\right|_{x=L/2}<0,
\end{eqnarray*}
where $L=x_2-x_1$ is the distance between the kinks, and the negative
sign of the derivative corresponds to a positive quantity
asymptotically vanishing far to the right of the kink.  The obtained
sign corresponds to an attraction at large distances.  The attraction
is also expected for a pair of symmetric non-topological
solitons (in this case the same formula with an appropriate ${\bf m}$
applies).  Of 
course, for the case of a single-component order parameter ${\bf
  u}\equiv u$ this result is well known.  Even in a more
general case, we could expect to find the attraction between such
defects, as we already know that inhomogeneous configurations are
always thermodynamically unstable in the
system~(\ref{eq:most-general-cont}), (\ref{eq:charge-constraint}),
unless there are topological reasons for the stability.  The effect of
topological stability is also easy to understand here: equally charged
kinks (which are allowed, for example, if the potential $V(u)$ is
periodic) always repel.  In accord with Sec.~\ref{sec:scaling}, such
kinks would be pushed infinitely far apart 
unless
stabilized by the boundary conditions. 

A similar calculation can be repeated for any combination of
spatially separated defects, in arbitrary dimension.  In every case
the interaction in the lowest order can be split into a sum of
pairwise terms which are defined by the gradient terms in the original
free energy.  

\section{Symmetry and the structure of domain walls}
\label{sec:no-twist}

So far we mostly considered global properties of the configurations
minimizing the free energy of the general
form~(\ref{eq:most-general-cont}).  For this {\em local\/} functional
we saw that non-uniform states are unstable to phase separation, and
thus indicated the Coulomb repulsion as an important component of any
continuous mean-field model designed to describe the observed
incommensurate structures in high-$T_c$ materials.  Now let us
concentrate on the {\em local\/} structure of non-uniform
configurations minimizing the free
energy~(\ref{eq:most-general-cont}),
(\ref{eq:long-distance-interaction}).  Specifically, we shall attempt
to answer the question whether a component of the order parameter can
change its sign in a thermodynamically stable state (ground state
configuration).

For this question to make sense, the zero value must have an
unambiguous meaning.  This is guaranteed if the free energy 
depends only on the square 
of the order parameter.  
For example, in antiferromagnets time reversal symmetry assures
that this is the case for the pseudovector of magnetization
${\bf s}$.  Even if the full spin-rotational symmetry is broken, 
 the susceptibilities $\chi_i$, the
potential $V$, and the charge density $\rho$ can only depend on the
bilinear combinations $s_i\,s_j$ of the magnetization components.  The
free energy will depend only on the squares $s_i^2$ as long as the
mixed combinations with $i\neq j$ are prohibited by the symmetry, as
discussed in Sec.~\ref{sec:count-groups}.

\subsection{Continuous symmetry and the untwisting instability}
\label{sec:cont-symmetry}

Let us first consider a system with a free energy of the
form~(\ref{eq:most-general-cont}),
(\ref{eq:long-distance-interaction}), with an additional rotational
symmetry between $m\ge 2$ components of the order parameter ${\bf
  u}=(s_1,\ldots,s_m,\phi_1,\ldots)$.  For clarity, and having in mind
a particular application to magnets, we shall call these the components 
of a
(generalized) spin magnetization ${\bf s}$, and assume that both local
and non-local parts of the free energy can only depend {\em
  analytically\/} on the square $S^2\equiv {\bf s}^2$ of this vector,
while the dependence on the remaining components $\phi_i$ remains
generic,
$$
\rho({\bf u})\equiv \rho({\bf s}^2,\phi_1, \ldots),\;
V({\bf u})\equiv V({\bf s}^2,\phi_1, \ldots), \;\ldots
$$
In the presence of such {\em continuous\/} spin-rotational
symmetry, the gradient terms in the free
energy~(\ref{eq:most-general-cont}) tend to align the direction of the
magnetization ${\bf s}$.  Indeed, the rotationally-symmetric gradient
term can be written as
\begin{equation}
  \label{eq:gradient-energy-decomposition}
  \chi_s(S^2,{\phi}_i)\,(\nabla {\bf s})^2=
    \chi_s(S^2,{\phi}_i)\,\left[(\nabla S)^2 
  + S^2(\nabla \hat {\bf e})^2\right],
\end{equation}
where $\hat{\bf e}\equiv {\bf s}/S$ is a unit vector in the
direction of ${\bf s}$.  Obviously, in any region where $S\neq 0$, the
energy of a ``twisted'' configuration ($\hat{\bf e}\neq{\rm Const}$)
can be lowered by aligning the magnetization along a common
direction, which eliminates the second term in the r.h.s.\ of
Eq.~(\ref{eq:gradient-energy-decomposition}).  The rotational
stiffness vanishes if $S=0$ (nodal points in one-dimensional case, or
nodal hypersurfaces for $D>1$), and the energy does not depend on the
relative orientation of the vectors ${\bf s}$ in the regions separated
by such nodes.  In any case, one can select $s_1=\pm S$, $s_l=0$ for
$l>1$, {\em i.e.}, the minimal configuration can be always chosen to
have only one component, although the sign of this component is not
fixed at this point.  We shall show below, however, that the energy of any
such configuration with a node (closed nodal surface for $D>1$) can be
continuously lowered by introducing an appropriately chosen
perturbation in the orthogonal direction.  Such instability to local
``untwisting'' is well-known for one-dimensional systems; it implies
that only uniformly-oriented spin configurations can minimize the free
energy in the presence of a rotational symmetry.

To analyze the ``untwisting'' instability in general, consider a spin
configuration ${\bf s}=(s_0,0)$ with a single non-zero component
$s_0(x)$ which is presumed to have a node (nodal surface for $D>1$).
The local instability of such configurations can be demonstrated by
introducing an orthogonal perturbation ${\bf s}_1=(0,s_1)$.  The
relevant part of the perturbed free energy
functional~(\ref{eq:most-general-cont}) can be written as
\begin{equation}
  \label{eq:spin-rotat-part}
 {\cal F}=\int d^Dx \left\{\chi(S^2,x)\,\left[(\nabla s_0)^2+(\nabla
 s_1)^2\right] +V(S^2,x)\right\},
\end{equation}
where $S^2=s_0^2+s_1^2$, and the additional coordinate dependence is
introduced to account for a possible presence of the remaining
non-uniform components of the order parameter.  Here we only consider
a simpler case in which  the charge density $\rho$ (and, consequently, the
long-range Coulomb interaction) 
is
independent of the spin configuration; this is generalized in
Appendix~\ref{app:untwisting}.

To quadratic order in the perturbation $s_1$ the increment of the free
energy~(\ref{eq:spin-rotat-part}) is
$$
\delta{\cal F}=\int \left\{\chi_0(x)\,(\nabla
  s_1)^2+G_0(x)\,s_1^2\right\} \,d^Dx,
$$
where the effective susceptibility $\chi_0(x)\equiv
\chi(s_0^2,x)>0$ is positive everywhere, the effective potential
$G_0(x)\equiv \chi'(s_0^2,x)\,(\nabla s_0)^2+V'(s_0^2,x)$ is
continuous and limited from below, and primes denote derivatives with
respect to $S^2$.  The local stability of the configuration $s_0(x)$
requires that the functional $\delta{\cal F}$ is
non-negative; equivalently, the self-adjoint eigenvalue
problem
\begin{equation}
  \label{eq:self-adjoint-ev}
  -\nabla(\chi_0(x)\,\nabla
  \varphi)+G_0(x) \varphi =\Lambda\varphi
\end{equation}
should have no negative eigenvalues.  Using the spin-rotational
symmetry (or directly, by comparing with the Euler-Lagrange equation
for $s_0$), it is easy to see that the function $\varphi_0(x)\equiv
{\rm Const}\,s_0(x)$ satisfies Eq.~(\ref{eq:self-adjoint-ev}) with
zero eigenvalue $\Lambda_0=0$.  It is a well-known fact about the
self-conjugate eigenvalue problem~(\ref{eq:self-adjoint-ev}) that its
ground state is non-degenerate and does not change
sign\cite{CourantHilbert}.  Since the function $\varphi_0(x)$ does
change its sign by assumption, it cannot be the ground state
eigenfunction, and, therefore, there must be at least one unstable
direction $\varphi_{-1}(x)$ which corresponds to a lower eigenvalue
$\Lambda_{-1}<\Lambda_0=0$.  Therefore, the energy of the original
spin configuration $(s_0(x),0)$ can be continuously lowered by the
orthogonal perturbation ${\bf s}_1={\rm Const}\, (0,\varphi_{-1}(x))$,
and we conclude that {\em only a uniformly oriented spin configuration
  without nodes (nodal hypersurfaces for $D>1$) can realize the global
  minimum\/} of the functional~(\ref{eq:most-general-cont}) in the
presence of a continuous spin-rotation symmetry\cite{endn:four}.

\subsection{Instability in the Ising limit}
\label{sec:discrete-symmetry}

Let us now imagine that the continuous spin-rotational symmetry is
broken by the lattice.  We begin with the case of a relatively strong
easy-axis (Ising) anisotropy, so that effectively only one component
$s$ of the spin remains.  In the absence of any other magnetic
ordering, the residual symmetry of the free energy is the discrete
$Z_2$ group associated with the time-reversal symmetry $s\to-s$.
Ordinarily, such broken symmetry indicates the possibility of {\em
  topologically stable\/} kinks, or domain walls in $D>1$, separating
regions of opposite magnetization.  
It turns out, however, that despite their local stability, such
configurations do not occur in the lowest energy state of
the system; they can only occur as excitations.  Formally, this can
be proven in general, utilizing the residual symmetry $Z_2$ of the
free energy functional.

Indeed, we saw that in the presence of a continuous spin-rotational
symmetry the ground state configuration is uniformly aligned; it can
always be chosen to have only one component of the spin.  Therefore,
the ground state of the functional
$$
{\cal F}=\int d^Dx \left\{\chi(s^2,x)\,(\nabla s)^2
  +V(s^2,x)\right\}, 
$$
is in a one-to-one correspondence (modulo the overall rotation)
with the ground state of the $U(1)$-symmetric extended functional
$$
{\cal F}_{\bf z}= \int d^Dx \left\{\chi({\bf s}^2,x)\,(\nabla
  {\bf s})^2 +V({\bf s}^2,x)\right\},
$$
where the field ${\bf s}=(s_{1},s_{2})$ has two components.
Because of the untwisting instability the second functional has a
nodeless ground-state configuration; our mapping indicates that so
does the first. 

We have proven a version of the no-node theorem, {\em i.e.\/}\ the
statement that any component $s$ of the order parameter {\em preserves
  its sign in the globally minimal configuration}, provided that the
potential energy (including the long-distance part, see
Appendix~\ref{app:untwisting}) depends only on the square of this
component.\cite{endn:five}

\subsection{Group-theoretical analysis:  Effects of ``Spin Orbit 
Coupling''}
\label{sec:count-groups}

The situation of perfect Ising anisotropy considered in the previous
section is, of course, an idealized case.  In real systems the
anisotropy can be quite small, so that all three components
$(s_x,s_y,s_z)$ of the magnetization pseudovector must be considered.
Nevertheless, it is possible to show that the same conclusion about
the absence of topological domain walls holds as long as the symmetry
of the underlying lattice is high enough.

Generally, because of the global time-reversal symmetry, the local
potential energy can be an arbitrary function of all bilinear
combinations $s_i\,s_j$, $i,j=x,y,z$.  Expanding in powers of such
products, we can also write any such function as
\begin{equation}
  \label{eq:group-expansion}
  V(s_i\,s_j) =V_0+V_1\,s_y\,s_z+V_2\,s_z\,s_x+V_3\,s_x\,s_y,
\end{equation}
where the coefficients in the expansion are, generally, some
functions of the squares of the magnetization components, $V_k\equiv
V_k(s_x^2,s_y^2,s_z^2)$, $k=0,\ldots,3$.  The statement about the sign
of the magnetization components proven in the previous section applies
only if the cross-terms are absent.  In particular, this happens
independently of the specific details of the function $V(s_i\,s_j)$,
if such terms are not allowed by the symmetry of the lattice.
Conversely, if at least one of such terms {\em is\/} present, no
general statement about the sign of any component of the spin
magnetization can be made, unless the additional components of
magnetization are suppressed by a sufficiently strong easy axis
anisotropy.

The effective free energy functional should remain invariant under any
transformation which preserves the lattice structure; for the local
potential $V$ only the transformations from the corresponding
crystallographic point group are relevant.
Because the pseudovector of magnetization remains invariant under
inversion, its components transform under reflection
$$
\sigma_h:\quad (x,y,z)\to (x,y,-z)
$$
as $(s_x,s_y,s_z)\to (-s_x,-s_y,s_z)$, in exactly the same fashion
as under the $\pi$-rotation with respect to the axis $z$,
$$
C_2:\quad (x,y,z)\to (-x,-y,z).
$$
The invariance of the potential~(\ref{eq:group-expansion}) with
respect to either of these transformations requires $V_1=V_2=0$.  The
existence of another symmetry transformation of one of these kinds,
with respect to an orthogonal plane or an orthogonal axis, is
sufficient to suppress the only remaining coefficient, $V_3=0$.

Such symmetries are present in all crystallographic point groups of
cubic (groups $O$, $O_h$, $T$, $T_h$, $T_d$) and orthorhombic
($C_{2v}$, $D_2$, $D_{2h}$) systems, and in sufficiently symmetric
groups of tetragonal ($C_{4v}$, $D_4$, $D_{4h}$, $D_{2d}$) and
hexagonal ($C_{6v}$, $D_6$, $D_{6h}$, $D_{3h}$) systems.  For all
other crystallographic groups we constructed invariant expressions
mixing several components of the magnetization.  For example, the
quantity $ s_x \,s_y\,(s_x^2-s_y^2)$ is symmetric with respect to all
transformations of the groups $C_4$, $C_{4h}$ and $S_4$, the quantity
$s_z\,s_y\,(s_y^2-3s_x^2)$ is symmetric with respect to all trigonal
groups,
{\em etc}.

The lattice symmetry also determines the structure of the derivative
terms in the free energy functional.  In addition to components of the
pseudovector of the magnetization ${\bf s}$, we now have the
components of the axial vector of the gradients, and so the number of
possible symmetric terms increases.  The conclusions about the phase
separation and the local structure of the domain walls will be
absolutely modified if the terms {\em linear\/} in derivatives are
present in the free energy.  Such terms are known to stabilize
topological domain walls in the ground state.  Among the groups we
listed above, only the groups $O_h$, $T_h$, $D_{6h}$, $D_{4h}$, and
$D_{2h}$ absolutely prohibit the existence of invariant quantities
linear in derivatives.  All these groups include the inversion, which
guarantees the absence of such invariants.  The groups which include
only proper rotations were eliminated by the existence of the
pseudoscalar invariant ${\bf s}\cdot [\nabla\times{\bf s}]$.  All
other groups required special consideration.\cite{endn:six}

The highly-symmetric point groups listed in the previous paragraph
prohibit both linear in the derivatives terms, and the mixing between
different components of the magnetization in the potential energy.
Nevertheless, in the presence of spin-orbit interaction {\em any\/}
point symmetry group allows mixing between different components of the
magnetization in the gradient terms due to the existence of a
rotationally-invariant scalar
$$
(\nabla\cdot {\bf s})^2=
  (\nabla {s}_x)^2+2\, \nabla {s}_x\,\nabla s_y+\ldots
$$
For specific groups, dangerous terms can also include
less-symmetric invariant quantities containing the terms of the form
$\partial_x s_x\,\partial_y s_y$.  Formally, because these terms
cannot be eliminated by symmetry, antiphase domain walls are possible
in the ground state of any non-Heisenberg system.  For the
case of magnetic ordering one may argue, however, that the symmetry
breaking in the gradient terms can only result from the combination of
the hopping, already small because it is determined by the tunneling
matrix elements, and the spin-orbit interaction, typically small
because it is a relativistic effect.  Therefore, such terms are
expected to be very small, and it is clear that they cannot be
responsible for very robust antiphase domain wall ordering observed in
the cuprates and nickelates.

\section{Antiphase domain walls}
\label{sec:antiphase}

The crystallographic point groups of the relevant phases of high-$T_c$
materials\cite{Cox-87,Cox-88,Axe-89B,Axe-89A,Cox-89} and related
com\-pounds\cite{Yimei-94,Moodenbaugh-98} are $D_{4h}$ in tetragonal
phases, and $D_{2h}$ in orthorhombic phases.  According to our
arguments in the previous section, these highly-symmetric groups
absolutely rule out antiphase domain walls in the lowest energy state,
and yet such domain walls have been observed in many such materials.
Moreover, this constraint is not limited to the continuous
model~(\ref{eq:most-general-cont}) with quadratic in the derivatives
gradient terms: many lattice models with arbitrary long-distance
interactions can be cast in the generic form considered in
Appendix~\ref{app:discrete}, and by the theorem proven there they must
have ground states with uniform sign of the order parameter.  Clearly,
this situation is by no means an exception!

In the remaining part of the paper we show that
antiphase domain walls in the ground state can be stabilized in
the presence of  {\em frustration\/} involving
competing interactions.  We consider two specific models with
short-range interactions:
a lattice model of a doped antiferromagnet, and a continuous model
with higher-order derivative terms.  In both systems periodic antiphase
domain wall structures can be thermodynamically stable at large enough
charge densities, but domain walls attract at asymptotically large
distances, so that the phase separation necessarily happens at
sufficiently small values of doping.

\subsection{Antiphase domain walls on the lattice}
\label{sec:lattice}
\noindent

Let us consider a lattice model of the form
\begin{equation}
  \label{eq:lattice-af}
  {\cal F}=J\sum_{\langle ij\rangle} {\bf S}_i{\bf S}_j
  +J'\sum_{\langle ilj\rangle}  {\bf S}_i n_l {\bf S}_j
  +\sum_i V({\bf S}_i^2,n_i),
\end{equation}
where the first term represents the usual exchange of localized spins,
the second term\cite{Castro-96,Zaanen-98} is due to higher order
exchange processes with virtual hops through a partially occupied
site, the hole density, $0\le n_l\le1$ is defined to be a bounded
continuous variable, and the local potential $V$ must be chosen to
insure the stability of the model, as well as to provide an adequate
repulsion between the holes and the spins on the same site.  As usual,
we presume that the average hole density is  fixed,
\begin{equation}
  x\equiv \bar n={\cal N}^{-1}\sum_i n_i,
  \label{eq:lattice-constraint}
\end{equation}
where ${\cal N}$ is the total number of lattice sites.  Clearly, the
positive values of the second exchange constant $J'>0$, tend to
frustrate antiferromagnetic ordering in a doped system; we argue below
that a competition of this sort is necessary to form antiphase domain
walls and suppress the global AF order in the system. 

For the purpose of this example, we will limit our analysis to the
quartic form of the potential
\begin{equation}
  \label{eq:lattice-pot}
  V({\bf S}^2, n)={g_1\over2}({\bf S}^2-1)^2+\left(\tilde g_2 n+{z
  J\over 2}\right)\,{\bf S}^2 
  +{g_3\over2} n^2, 
\end{equation}
where $\tilde g_2=g_2-z(z-1) J'/2$, $z$ is the lattice coordination
number, and the coefficients are chosen so that in terms of the
antiferromagnetic Ne\`el order parameter ${\bf s}_i=(-1)^i{\bf S}_i$
the free energy could be rewritten in a form
\begin{eqnarray}
  \label{eq:lattice-neel}
{\cal F}&=&{J\over2}\sum_{\langle ij\rangle} ({\bf s}_i-{\bf s}_j)^2
  +J'\sum_{\langle ilj\rangle}\, n_l({\bf s}_i{\bf s}_j-{\bf s}_l^2)
  \nonumber\\
  &\relax&+\sum_i\left[ 
  {g_1\over2}({\bf s}_i^2-1)^2+ g_2\, n_i\,{\bf s}_i^2  +{g_3\over2}
  \,n_i^2\right].  
\end{eqnarray}
The term with the coefficient $g_1$ favors unit values of the on-site
magnetization, the coefficient $g_2$ is a measure of the strength of
the repulsion between spins and charges, while the coefficient $g_3$
measures the local tendency against doping.  

At zero doping all charges necessarily vanish, $n_l=0$, and
Eq.~(\ref{eq:lattice-neel}) is minimized by a uniform AF state ${\bf
  s}^2=1$ with the value ${\cal F}_{\rm AF}(0)=0$.  Uniform AF
states can be also formally found at sufficiently small non-zero
dopings, with energy given by the second line of
Eq.~(\ref{eq:lattice-neel}), minimized at ${\bf s}^2=1-g_2\,x/g_1\ge0$
with the energy density value
$$
{f}_{\rm AF}(x)=g_2\,x+{x^2\over2}\left(g_3-{g_2^2\over g_1}\right). 
$$
The magnitude of the AF ordering reduces to zero at $x=g_1/g_2$,
and at larger filling fractions the AF phase is replaced by a uniform
non-magnetic state with the energy $f_0=(g_1+g_3 \,x^2)/2$. 

\begin{figure}[htbp]
  \epsfxsize=\columnwidth%
  \epsfbox{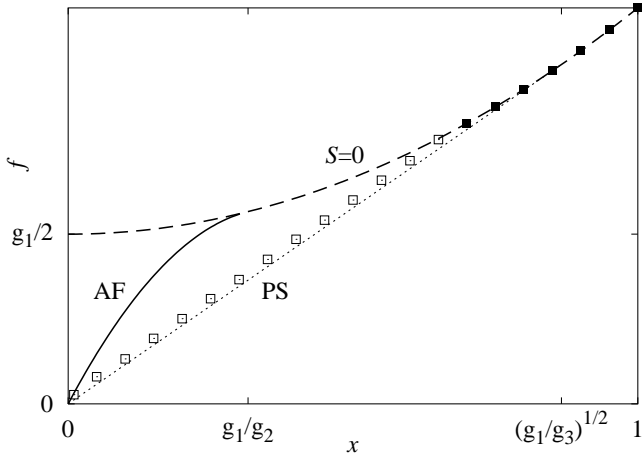}
  \caption{Locating the minimum of the free
    energy~(\protect\ref{eq:lattice-neel}) per unit site for the
    strong repulsion case, 
    $J=0.9$, $J'=0.6$, $g_1=0.6$, $g_2=1.9$, $g_3=0.8$.  Bold solid
    and dashed lines respectively show the energies of uniform AF and
    non-magnetic ($S=0$) phases.  Dotted line gives the free energy
    per site of an
    infinite system in the phase-separated regime.  Solid and empty
    squares respectively indicate periodic and phase-separated 
    configurations minimized numerically with system sizes up to ${\cal
      N}=40$.}
  \label{fig:energ1}
\end{figure}

\begin{figure}[htbp]
  \epsfxsize=\columnwidth%
  \epsfbox{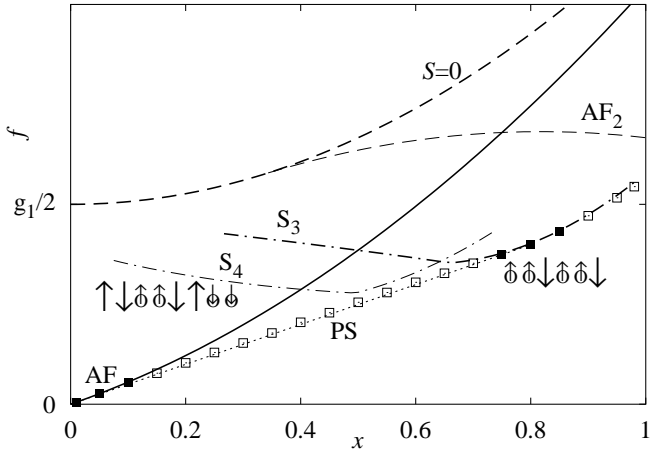}
  \caption{Locating the minimum of the free
    energy~(\protect\ref{eq:lattice-neel}) per unit site for the case
    of weak repulsion, $J=0.9$, $J'=0.6$, $g_1=0.6$, $g_2=0.3$,
    $g_3=0.8$.  The line AF$_2$ corresponds to a uniform AF with the
    period of four lattice sites, which becomes preferable at larger
    values of $J'$.  The lines S$_3$ and S$_4$ correspond to
    commensurate stripe phases with the charge periods $3$ and $4$, as
    illustrated in the insets.  Below $x\approx 0.75$ the system phase
    separates into an undoped (or very weakly doped) AF phase and the
    phase S$_3$.  Solid and empty squares respectively indicate the
    phase-separated and uniform configurations as seen numerically
    with system sizes up to ${\cal N}=40$.}
  \label{fig:energ2}
\end{figure}

The energies of these phases for the strong repulsion case
$g_2^2>g_1\,g_3$ are illustrated in Fig.~\ref{fig:energ1}.  The
function ${f}_{\rm AF}(x)$ (solid line) has a negative curvature
at small values of 
doping, so the system is necessarily unstable to phase separation
between an undoped antiferromagnet and a completely or partially doped
uniform non-magnetic phase (dashed line).  The energy of
phase-separated system is shown in 
Fig.~\ref{fig:energ1} with a dotted line. 
The absence of other phases was checked
numerically by minimizing Eq.~(\ref{eq:lattice-neel}) for systems with
periodic boundary conditions of all even sizes in the range between
${\cal N}=4$ and ${\cal N}=40$.  To reduce the possibility of
accidental trapping in a local minimum, we used the Metropolis
algorithm with variable temperature (simulated annealing).  For each
system size we did a set of up to 8 trial cooldown runs starting with
a random configuration, selected the best resulting configuration, and
then repeatedly cycled the temperature up to 20 times.  The minimal
energy density chosen among the systems of all sizes was used as an
estimate of the ground state energy; these values are shown in
Figs.~\ref{fig:energ1}--\ref{fig:energ2} with squares.  As expected, in
the regime of phase separation, typically the lowest energy density
was achieved for the biggest system.

Phase separation is impossible if a long-distance interaction is also
included in the model (\ref{eq:lattice-af}).  However,  the above
calculation 
remains relevant as long as this interaction is sufficiently
weak.  In this case, there 
exists a large length scale $D$, at which the long distance forces
become relevant.  It is this scale that determines the period of a
stripe phase, in which the regions of undoped AF and non-magnetic
phase are separated by the domain walls of the model
(\ref{eq:lattice-af}).  As long as the size $d$ of these domain walls
is relatively small, $d\ll D$, the long-range interaction does not
significantly change their form.

In the considered regime of the strong local repulsion, $g_2^2\gg
g_1\,g_3$, the domain wall between the undoped AF and the non-magnetic
phase with the density $x=\min[1,\,(g_1/g_3)^{1/2}]$ is very sharp.
The order parameters approach their vacuum values as determined by the
solution of the corresponding linearized equations.
On the AF side,
the charge density is locked at $n=0$, and the perturbation $\delta
{\bf s}_j\sim \exp(-\kappa_0 j)$ falls off with the same exponent as
in the ideal undoped AF,
\begin{equation}
  \label{eq:lattice-undoped-decay}
  \sinh^2(\kappa_0/2)=g_1/J. 
\end{equation}
Similarly, expanding the free energy~(\ref{eq:lattice-neel}) to
quadratic order in the vicinity of the zero-magnetization state with the
density $n_1=(g_1/g_3)^{1/2}<1$, we obtain
\begin{eqnarray*}
4\sinh^2\left(\kappa_1/2\right)&=&
\left(2-{J\over 2 n_1 J'}\right)\\
& \relax& 
+\sqrt{\left(2-{J\over 2 n_1 J'}\right)^2+{2(g_2 n_1-g_1)\over n_1
    J'}}. 
\end{eqnarray*}
The second term under the square root, and, consequently, the r.h.s.\ 
of the entire expression, are guaranteed to remain positive everywhere
in the strong repulsion regime, independent of the values of the
exchange constants.  The domain walls are relatively narrow when
$\kappa D \ll 1$; in this case the solution has a form of an
array of domain 
walls between the AF and nonmagnetic regions.  This is the canonical
picture of Coulomb-frustrated phase separation\cite{Emery-93,Low-94}, where
wide stripes are directly analogous to the classical stripe
phases\cite{Andelman}. 

The ground state phase diagram changes substantially in the opposite
case of very weak repulsion, $g_2^2\ll g_1\,g_3$.  The main difference
of this regime is that nonuniform phases with anti-phase domain walls
are much closer to stability; as illustrated in Fig.~\ref{fig:energ2},
some of them may be stable even in the absence of any long-range
forces.  As the long-range interactions are introduced, instead of
stabilizing wide stripes by the usual Coulomb-frustrated phase
separation\cite{Emery-93,Low-94} mechanism, they may stabilize certain
dense stripe phases.  Such picture of Coulomb-stabilized microscopic
stripe phases is in agreement with the arguments of Hellberg and
Manousakis\cite{Hellberg-99} based on their results of exact numerical
diagonalization of small $t-J$ clusters.
 
In the considered limit of weak repulsion, $g_2^2\ll g_1\,g_3$,
non-zero magnetization can coexist with substantial doping even in the
limit of a fully doped system $x=1$.  Because of the constraint $0\le
n_i\le1$, only a uniform charge configuration is possible at $x=1$,
and the spin ordering is determined by the competition between two
exchange couplings.
For a particular set of parameters chosen in Fig.~\ref{fig:energ2},
the lowest-energy phase in this limit has a spin modulation period of
three lattice sites.  As the doping is reduced, it is energetically
favorable to put all electrons at the points of maximum magnetization,
so that the charge density has a period of three lattice sites, as
illustrated in the right caption.  The energy of such a {\em
  ferrimagnetic\/} phase S$_3$ is denoted with a bold dash-dotted line
in Fig.~\ref{fig:energ2}; as the doping is lowered, this line starts
to increase again below the point $x\approx 0.67$ where single undoped
sites are separated by fully doped antiphase domain walls of width two
sites.  In a similar phase S$_4$ (with the charge period of four and
the spin period of eight sites), such domain walls are separated by
two weakly doped sites, but this phase is avoided in large systems
which prefer to phase separate instead.  The energy density of
phase-separated system (PS) is shown with the dotted line; in the
vicinity of the point $x\sim 0.5$ this line goes only slightly below
the line denoting the energy of the stripe phase $S_4$.

Numerically, for all combinations of parameters we tried, the
non-uniform ``stripe'' phases seemed to be stable only at
sufficiently large values of doping.  It turns out that this statement
can be proven for any form of the potential $V({\bf s}^2,n)$ in
Eq.~(\ref{eq:lattice-af}) by using a variant of the argument in
Sec.~\ref{sec:interaction}.  Any non-uniform charge configuration in
the limit of low doping must consist of some defects, charged solitons
or domain walls, separated by wide regions of almost perfect AF.  In
this limit every defect, described by the spin ${\bf s}_i$ and
charge $n_i$ distributions, must realize a local minimum of the free
energy~(\ref{eq:lattice-af}), and satisfy appropriate
Euler-Lagrange equations.  A two-defect configuration can be well
approximated by a linear superposition of corresponding spin- and
charge-density distributions, with the value of the
constraint~(\ref{eq:lattice-constraint}) independent of the mutual
position of the defects.  In the vicinity of each defect the effect of
the other one can be considered as a perturbation.  By rearranging the
sums independently in each region, with the help of the corresponding
Euler-Lagrange equations, the linear order cross-terms can be made to
disappear in the bulk, so that only the ``integrated'' part
\begin{eqnarray}
  \nonumber
  \delta E&=&+J\,\delta{s}_0^{b}\,\delta{s}_{1}^{a}
  +J'\left[{s}_{-1}^{a}\,n_0^{a}\,\delta{s}_1^{b}
      - \delta{s}_0^{b}\,n_1^{a}\,{s}_2^{a}\right]\\
  &\relax &- J\,\delta{s}_0^{a}\,\delta{s}_{1}^{b}
  -J'\left[{s}_{-1}^{b}\,n_0^{b}\,\delta{s}_1^{a}
      - \delta{s}_0^{a}\,n_1^{b}\,{s}_2^{b}\right]
  \label{eq:lattice-interact}
\end{eqnarray}
remains.  Here $\delta s\equiv s-s_\infty$ is the deviation of the AF
magnetization from its vacuum value, and the superscripts $a$ and $b$
label the fields caused by the defect situated far to the left and
far to the right from the origin, respectively.  Similarly to
Eq.~(\ref{eq:interaction-energy-simple}), the precise location of the
separation boundary is not important, as long as it is chosen far
enough from each defect.  For a symmetric defect configuration,
$s^a_l=s^b_{1-l}$, Eq.~(\ref{eq:lattice-interact}) can be rewritten as
\begin{eqnarray}
  \delta E&=&J\left[(\delta{s}_{1}^{a})^2-(\delta{s}_{0}^{a})^2\right]
      +2 J'\,{s}_{\infty}^{a}\left[ 
          \delta n_0^{a}\,\delta{s}_0^{a}-\delta
      n_1^{a}\,\delta{s}_1^{a}\right]\nonumber\\ 
  &\relax&
      +2 J'\,n_\infty^{a}\left[\delta{s}_{-1}^{a}\,\delta{s}_0^{a}
      - \delta{s}_1^{a}\,\delta {s}_2^{a}\right],
  \label{eq:lattice-interact-reduced}
\end{eqnarray}
where $\delta n_l\equiv n_l-n_\infty$.  Only the first term exists for
the asymptotic form~(\ref{eq:lattice-undoped-decay}), where the hole
density $n_l$ is pinned to zero at finite distances from defects.
This term gives a {\em negative\/} interaction energy, corresponding
to asymptotic {\em attraction\/} between far separated defects.  This
is in accordance with our simulation in Fig.~\ref{fig:energ2}, where
the most stable charge-modulated configuration was a dense condensate
of antiphase stripes.  Of course, the repulsion of the stripes at
small distances, and the stability of the dense stripe configuration
cannot be inferred from this asymptotic analysis.

Generally, for models of the form~(\ref{eq:lattice-af}), the hole
density $n_l$ does not necessarily vanish at a finite distance from a
defect, or it may even have a non-zero value $n_\infty$ in the
intermediate AF phase.  Then the second exchange term also contributes
to the interaction energy.  In principle, this contribution may be
attractive or repulsive, depending on the relative sign of
$s_\infty\delta s$ and $\delta n$.  However, we are interested in
systems with strong repulsion between AF ordering and the doped holes;
here the effect of the second exchange is negative, and the second
term in the first line of Eq.~(\ref{eq:lattice-interact-reduced}) gives
attraction as well.

Contrarily, the {\em last\/} term in
Eq.~(\ref{eq:lattice-interact-reduced}), which exists only if the
doping saturates to a non-zero value $n_\infty$ far from the solitons,
is positive, it contributes to a repulsion between the domain walls.
This is not surprising, because the second exchange term counteracts
the usual exchange if a finite hole density is present.  Nevertheless,
one can show that the net result is an attraction between the defects,
as long as the 
uniformly doped AF state is locally stable.  

\subsection{Antiphase domain walls in a continuum model}
\label{sec:quartic}

Although we now have an example of a model which admits antiphase
domain walls in the ground state, this model is not a continuum model,
and one might infer that it is the lattice commensuration effects that
enable the existence of antiphase domain walls in the ground state.
To stress our statement that it is not the lattice, but the
frustration between different interactions that stabilizes such domain
walls, we give a brief analysis of a continuum model with similar
properties.

Consider a one-dimensional system with the free energy of the form
\begin{eqnarray}
  {\cal F}&=&\int dx\biggl[
  \beta\, ({\bf s}'')^2+\chi_s\,({\bf s}')^2+\chi_\phi\,(\phi')^2
    \label{eq:high-deriv-fe}
    +V({\bf s}^2,\phi)
  \biggr]
\end{eqnarray}
As usual, the primes denote spatial derivatives, the field ${\bf s}$
represents an antiferromagnetic order parameter, and $\phi$ is a
scalar field with some conserved charge density $\rho=\rho(\phi)$.
Unlike Eq.~(\ref{eq:most-general-cont}), we no longer assume that the
spin susceptibility 
$\chi_s=\chi_s(\phi)$ 
is a positively-defined
function of the scalar order parameter $\phi$, and the higher-order
derivative term, with $\beta >0$, is required for stability.  In
analogy with the second 
hopping term of the lattice model~(\ref{eq:lattice-af}), we shall
assume that the spin susceptibility
\begin{equation}
  \label{eq:spin-suscept}
  \chi_s(\phi)=1-\alpha\,\rho(\phi)
\end{equation}
depends linearly on the charge
density, so that its sign can be reversed in the presence of large
enough hole density.

\noindent{\bf Scaling analysis}:  
It is obvious that the general conclusion of instability of periodic
states made in Sec.~\ref{sec:scaling} does not apply for the
model~(\ref{eq:high-deriv-fe}).  Indeed, instead of
Eq.~(\ref{eq:short-range-scaling}), we obtain 
\begin{equation}
  \label{eq:quartic-scaling}
  f_\lambda=\lambda^4\,{\rm Q}_1+\lambda^2\,{\rm K}_1+\Pi_1,  
\end{equation}
where ${\rm Q}_1>0$ is the contribution of the term(s) quartic in the
derivatives.  Because the second derivative terms are no longer
positively defined, this expression may have a minimum at $\lambda=1$
and
$$
K_1=-2Q_1<0. 
$$
Although this condition does not {\em guarantee\/} the global
stability of a periodic solution, it is clear that periodic structures
{\em may\/} in principle be stabilized for the free
energy~(\ref{eq:quartic-scaling}).

\noindent{\bf Asymptotic interaction of domain walls}:
\noindent 
The asymptotic form of the interaction between the domain walls for
the model~(\ref{eq:high-deriv-fe}) can be easily found by a linear
analysis similar to that in Sec.~\ref{sec:interaction}, by evaluating
the energy of a superposition of two domain walls separated by a wide
stretch of undoped antiferromagnet.  As before, only surface terms
survive in the linear order,
\begin{eqnarray*}
  \delta E&=& 
  2 \beta\,\delta{\bf s}_a''\delta{\bf s}_b'-2 (\beta\,\delta{\bf
    s}_a'')'\delta {\bf s}_b\\
  &\relax&+
  2\chi_s\,\delta{\bf s}_a'\delta{\bf
    s}_b+2\chi_\phi\,\phi_a'\phi_b-\bigl(a\leftrightarrow b\bigr),
\end{eqnarray*}
where the scalar field $\phi_i$ and the deviation of the AF order parameter
$\delta{\bf s}_i$ must satisfy the corresponding Euler-Lagrange
equations exactly, $i=a,\,b$ respectively denotes the defect located
far to the left and far 
to the right of the point where this expression is evaluated.  For two 
symmetric domain walls ${\bf s}_a(x)={\bf s}_b(2x_0-x)$ this expression is
simplified if the point $x_0$ is chosen exactly in the middle,
$$
\delta E=\left.-4(\beta\,\delta{\bf s}_a'' \delta{\bf
    s}_a)'+2\chi_s\,(\delta{\bf s}_a^2)'
  +2\chi_\phi\,(\phi_a^2)'\right|_{x=x_0}.
$$
The parameters $\beta$, $\chi_s$ and $\chi_\phi$ in this expression
must be evaluated in the vacuum configuration; they are all positive.
The perturbation of the vacuum state gets smaller as we move to the
right, and the two 
last terms are negative; as before, this corresponds to an attractive
interaction.  However, it is easy to see that the first term is
positive; it contributes to repulsion between the domain walls.  Only by
analyzing the linearized Euler-Lagrange equations in
the nearly perfect AF region, we can conclude that the overall sign of
the interaction energy is negative, as long as the AF state is a
locally stable minimum of the functional~(\ref{eq:high-deriv-fe}).
Therefore, as previously, domain walls attract at large enough
distances, and the system cannot form a stable non-uniform solution at
asymptotically small doping as long as AF ground state is stable at
zero doping, and so long as there are no long-range forces.

\noindent {\bf Twist stability}:
The twist instability, which was discussed in
Sec.~\ref{sec:no-twist} for positive $\chi_s$ and $\beta=0$,
can be also avoided for the
model~(\ref{eq:high-deriv-fe}); a magnetization vector ${\bf s}$ can
reverse its direction and yet remain locally stable with respect to
twists.  A sufficient condition for this stability can be obtained by
analyzing the derivative terms in the free
energy~(\ref{eq:high-deriv-fe}).  By decomposing the vector ${\bf
  s}=S\,{\bf e}$ into a product of its magnitude $S$ and the unit
vector ${\bf e}$, after several integrations by parts, the
gradient terms in the free energy can be rendered into a form
\begin{eqnarray*}
  ({\bf s}'')^2&\to&
  S^2\,({\bf e}'')^2+( {\bf e}')^2\,
  \left[2  (S')^2-4 S'' S\right]  +( S'')^2, \\
  ({\bf s}')^2&=&( S')^2+S^2\,({\bf e}')^2. 
\end{eqnarray*}
The system~(\ref{eq:high-deriv-fe}) will remain stable to developing
spontaneous twists as long as the coefficient in front of $({\bf
  e}')^2$ remains positive; this gives the sufficient criterion of
stability, namely, the condition that the expression
\begin{equation}
  \label{eq:no-twist-criterion}
  2\beta\, ( S')^2-4 \beta S''\, S+\chi_s\,S^2
  >0 
\end{equation}
must remain positive everywhere.  This condition is easy to check
directly for any given single-component solution of Euler-Lagrange
equations; there is no need to look for multi-component solutions if
Eq.~(\ref{eq:no-twist-criterion}) is satisfied.

Formally, this expression can remain positive near a node of the
magnetization because of the presence of the higher-derivative term in
Eq.~(\ref{eq:high-deriv-fe}).  However, such solutions can be 
allowed in the ground state only if $\chi_s$ can become negative,
which indicates the presence of a competition between different
interactions.  Therefore, the role of the higher derivative term 
is only to limit the instability caused by this competition.

\noindent{\bf Approximate variational solution:}
To illustrate the considered general properties, let us choose the
potential 
\begin{equation}
  \label{eq:quartic-potential}
  V=  \int\left[
      {g_1\over2}({\bf s}^2-1)^2+ g_2\, \rho\,{\bf s}^2  +{g_3\over2}
      \,\rho^2\right]\,dx,  
\end{equation}
of the same quartic form as used in Eq.~(\ref{eq:lattice-neel}), with
$\rho\equiv \rho(\phi)=\phi^2$.  Numerically, the solutions at small
enough densities look very much like the usual domain walls in
magnets, with ${\bf s}$ changing its sign where $\phi$ has a maximum.
Although the simplest set of trial functions $\phi=\phi_0/\cosh(kx)$,
$s=\tanh(kx)$ does {\em not\/} work, we can use it as a variational
solution to estimate the ground state energy and the areas of
stability of different phases.

Performing the integration, we obtain the expression for the total charge
$$
{\cal Q}_0=\int dx \phi^2=2 \phi_0^2/k, 
$$
and the free energy
\begin{eqnarray*}
  {\cal F}_0& =&
  {2\over3k}
  \biggl[ {g_1} + {8\over5} {\beta } {k^4}
  +\, {k^2} \left( 2 + {\phi_0^2} - {8\over5} {\alpha } {\phi_0^2}
  \right)\\
  &\relax &\hskip1in + {\phi_0^2} \left( {g_2} + {g_3} {\phi_0^2}
  \right) \biggr] . 
\end{eqnarray*}
In the limit of small charge density the stripe solution must minimize
the energy per unit doped charge, $f_0\equiv {\cal F}_0/{\cal Q}_0$.
This is achieved by selecting the amplitude of the charge soliton
$$
\phi_0^4=({{g_1} + 2\,{k^2} + 8\,{\beta }\,{k^4}/5})/g_3. 
$$
The resulting expression has a minimum at a non-zero scale $k=k_0$ if
the constant $\alpha$ in Eq.~(\ref{eq:spin-suscept}) is
$$
\alpha ={5\over 8}\left[1+ {2\sqrt{g_3}\, \left( 1 +
      {{8\,{\beta }\,{k_0^2}}/{5}} \right) \over \sqrt{{g_1} +
      2\,{k_0^2} + {8\,{\beta }\,{k_0^4}}/{5}}} \right];
$$
the corresponding value of the energy per unit charge is
$$
f_0=  {g_2} + {\frac{2\,{\sqrt{{g_3}}}\,\left( {g_1} + {k_0^2} \right) }
  {{\sqrt{{g_1} + 2\,{k_0^2} + {{8\,{\beta }\,{k_0^4}}/{5}}}}}}
$$
The resulting configuration will be stable with respect to
twists if the criterion~(\ref{eq:no-twist-criterion}) is satisfied.
The analysis shows that this is indeed the case for large enough
values of $\beta$ and $g_3$.

The stability of a stripe phase made out of these domain walls is
determined by Eq.~(\ref{eq:quartic-scaling}).  With the derived
expressions we find
$$
{\rm K}_0=\left(1-{8\over 5}\alpha \right)+{2\over \phi_0^2}
=-{16\beta  k^2\over 5\phi_0^2}
$$ is always negative.  This implies that the periodic phase might
indeed be stabilized at some intermediate scale, in agreement with our
numerical simulations of this model.  Therefore, the local stability
of topological domain walls may lead to the stabilization of a {\em
dense\/} stripe phase made out of such walls, in agreement with
detailed simulations\cite{White-98A,White-98B} of the $t$-$J$ model.
However, such a phase can only be stable at large enough charge
densities: within the MF approximation we have shown that the
asymptotic large-distance interaction between such domain walls is
always attractive, and in the limit of small values of doping the
system necessarily phase separates.  In addition, more subtle
fluctuation effects\cite{PKH-98} always contribute 
to power-law Casimir attraction between charged defects, and the
statement about the phase separation in weakly-doped antiferromagnets
persists.

\section*{Conclusions.} 
Phase separation at small doping is a ubiquitous property of doped
insulators 
with short-distance interactions.  Generally, in the absence of a
frustration caused by competing interactions, the staggered magnetization of
the ground state never changes its sign.  These two statements can be
formulated as theorems in the vicinity of a second order phase
transition involving AF ordering, where the correlation length is
large, and the derivative terms are small.

In application to high-$T_c$ materials, the competition between the
tendency of the holes to move around, and the tendency of repulsive
interactions to localize the charges must be accounted for in any
model for describing high-T$_c$ superconductors or
related materials.  Only at relatively short distances (where,
strictly speaking, we go outside the limits of applicability of the MF
theory), the domain walls may repel, which could lead to the stabilization
of dense static stripe phases.

\noindent {\bf Acknowledgments:}  We would like to thank M.\ Fogler
and S.-C.\ Zhang for valuable discussions.  LPP was supported in part
by the grant DOE DE-FG02-90ER40542.  SAK was supported in part by the
NSF under grant DMR 98-08685 at UCLA.  VJE was supported by the
Department of Energy grant number DE-AC02-98CH10886.  YBB was
supported in part by the grant DMR 9814289 and under an IBM Research
Partnership Award.  EAD was supported by NSF at ITP.

\appendix
\section{Landau Theory}
\label{density-func}

The phase transition between a stripe phase and a high-temperature
disordered state considered by Zachar {\em et al}.\cite{Zachar-97}
involves only one spin order parameter, the incommensurate spin
density wave ${\bf S}_{{\bf q}}$.  The transition
{}from a well-developed antiferromagnet with a modulation vector
$\vec\pi=(\pi,\pi)$ to an incommensurate modulated phase must
account for both the original AF order parameter ${\bf S}_{\vec\pi}$ (which,
generally, cannot be assumed small), and the spin density wave ${\bf
  S}_{\vec\pi+\bf k}$, with modulation period $2\pi/k$.  Coupling
these two spin order parameters together, it is easy to write
non-trivial and yet spin-rotation-invariant terms of the Landau
expansion of the effective free energy,
\begin{equation}
\label{eq:cubic-simple}
{\cal F}=r_{\rm s}\,|{\bf S}_{\vec\pi+\bf k}|^2+r_{\rm c}\,|\rho_{\bf
  k}|^2 +\gamma\left[{\bf S}^*_{\vec\pi}\,{\bf S}_{\vec\pi+\bf
  k}\,\rho^*_{{\bf k}}+{\rm c.c.}\right]+\ldots, 
\end{equation}
where $\rho_{\bf k}$ is the complex-valued amplitude of the charge
density wave with the wave vector ${\bf k}$, $\rho^*_{\bf k}\equiv
\rho_{-{\bf k}}$, and the quartic (and higher order) terms required
for stability are omitted.  This expression suggests that an
instability in either the spin ($r_{\rm s}\equiv r_{\rm s}({\bf q})<0$) or the
charge ($r_{\rm c}\equiv r_{\rm c}({\bf k})<0$) sector generates both
spin- and charge-density waves at the wavevectors ${\bf
  q}=\vec\pi+{\bf k}$ and ${\bf k}$, respectively, with modulation
amplitudes linearly proportional to each other.  More precisely, the
modulation appears if $r_{\rm s}({\bf q})$ and/or $r_{\rm
  c}({\bf k})$ are negative, or if
$$
r_{\rm s}(\vec\pi+{\bf k})\,r_{\rm c}({\bf
  k})<|\gamma|^2\,|{\bf S}_{\vec\pi}|^2.
$$
Near the transition the magnitude of the incommensurate peak is
necessarily much smaller then the commensurate AF modulation $|{\bf
  S}_{{\vec\pi+\bf k}}|\ll |{\bf S}_{\vec\pi}|$; it is easy to see that this
corresponds to {\em in-phase\/} domain walls.  The derived
relationship between ${\bf q}$ and ${\bf k}$ implies that the periods
of spin and charge modulation must be equal for such domain walls.

Experimentally, novel incommensurate elastic peaks, coexisting with
the commensurate peaks at $(\pi,\pi)$, have been observed\cite{YoungLee-AF}
recently at the border of the antiferromagnetic region of
La$_{2-x}$Sr$_x$Cu\,O$_4$ at $x=0.05$.   The incommensurate
peaks are rotated by $45^\circ$ compared to the antiphase peaks at
larger doping,
which could be caused by the fact that these peaks appear at a
temperature that is lower than the energy of the LTO-LTT phase
mode.\cite{Wakimoto-private}
If the data represent a bulk effect,
and assuming that the commensurate AF correlation length in the
cluster spin-glass phase\cite{Niedermayer-98} at smaller
values of doping ($x<5\%$) is sufficiently large for the Landau
expansion~(\ref{eq:cubic-simple}) to apply, we interpret the
simultaneous presence of both commensurate and incommensurate peaks as
the signature of {\em in-phase\/} domain walls, expected in this
region, and not merely coexisting antiferromagnetic and stripe
phases.
The above analysis indicates that the corresponding charge modulation
must have the {\em same\/} period and direction as that of the SDW
order.  Because the observed ordering differs substantially for these
two phases, the transition from a weakly modulated diagonal AF phase
to the 
fully-developed stripe state with antiphase domain walls is expected
to be first order in clean system.

In general, however, a discontinuous transition between a topological
and non-topological stripe phases is not the only possibility.  A
particularly simple scenario of a continuous transition between these
phases corresponds to a {\em dimerization\/} transition, where pairs
of antiphase domain walls spontaneously merge to form wider dimerized
domain walls, similar to those observed in manganates.\cite{Mori-98}
As a result, the period of charge modulation doubles, and a CDW with
the periodicity of the original spin ordering must develop.  In
addition, the perfect symmetry between the regions with two opposite
signs of AF order is broken, and a net antiferromagnetic ordering
appears.  Here we present only the simplest scenario for such a continuous
transition, minimally extending the charge-driven part of the phase
diagram of Zachar {\em et al}.\cite{Zachar-97} A more complete symmetry
analysis of possible dimerized phases will be published
elsewhere.\cite{Bazaliy-99-Unpublished}

To describe the dimerization transition, the Landau effective free
energy must include at least two harmonics of the density wave,
$\rho_{\bf k}$, $\rho_{2{\bf k}}$, coupled to the SDW harmonics, ${\bf
  S}_{\vec\pi+l{\bf k}}$, $l=0,\,1,\,2$.  While the quadratic part of the free
energy has the usual form,
$$
{\cal F}_2=\sum_{l=0}^2 r_{s\,l}\left|{\bf S}_{\vec\pi+l{\bf k}}\right|^2 
+\sum_{l=1}^2 r_{\rho\,l}\left|{\rho}_{l{\bf k}}\right|^2 , 
$$
there are {\em five\/} possible cubic terms
\begin{eqnarray}
  \label{eq:cubic-dimer}
  {\cal F}_3&=&\rho^*_{2{\bf k}}\left(\lambda_0\,{\bf S}_{\vec\pi+{\bf
    k}}^2 
    +\lambda_1 \,\rho_{\bf k}^2+\lambda_2 \,{\bf S}_{\vec\pi}\,{\bf 
    S}_{\vec\pi+2{\bf k}}\right)\nonumber\\ 
    & & +\rho^*_{{\bf k}}\left(\gamma_0\,{\bf S}_{\vec\pi+{\bf
    k}}\,{\bf S}_{\vec\pi}
    +\gamma_1\,{\bf
    S}_{\vec\pi+2{\bf k}} \,{\bf S}^*_{\vec\pi+{\bf k}}\right)+{\rm
    c.c.}
\end{eqnarray}
The invariant with the coefficient $\lambda_0$ has been 
considered previously in Ref.~\CITE{Zachar-97}, and the 
terms with
coefficients $\gamma_0$ and $\lambda_2$ were considered above in
Eq.~(\ref{eq:cubic-simple}).

Let us follow Zachar {\em et al}.\cite{Zachar-97}\ and consider the
transition from a disordered phase, driven by the instability in the
CDW sector, $r_{\rho\,2}<0$.  In this scenario, as the
amplitude of the CDW $\rho_{2{\bf k}}$ gets sufficiently large, the
term with the coefficient $\lambda_0$ generates an instability in the
SDW sector.  
{}From our {\em extended\/} free energy~(\ref{eq:cubic-dimer}) it is
clear that the same density wave may also destabilize the
double-periodic CDW $\rho_{\bf k}$ (via the term with coefficient
$\lambda_1$).  If this is the case, the remaining cubic invariants
will simultaneously generate non-zero AF modulation ${\bf
  S}_{\vec\pi}$ (coefficient $\gamma_0$) and an additional SDW
harmonic ${\bf S}_{\vec\pi+2{\bf k}}$ (coefficients $\lambda_2$ and
$\gamma_1$).  Obviously, in a certain range of parameters, the
transition to the phase with $\rho_{\bf k}\neq 0$ is continuous.  The
resulting {\em dimerized\/} phase, with equal periods of SDW and CDW,
and a non-zero AF ordering, would be interpreted as a non-topological
stripe phase.  If observed, such transition will provide a precise
{\em macroscopic\/} meaning to the notion\cite{Mori-98} of dimerized
stripes.

\section{Interaction of charged solitons}
\label{app:charged-interact}

Here we demonstrate that the expression for interaction energy between
the defects, derived in Sec.\ \ref{sec:interaction}, also works for
systems with global charge constraint (\ref{eq:charge-constraint}).
The single-soliton field configuration ${\bf u}_{0}(Q;x)$ minimizes
the energy functional at a fixed value of charge $Q$, but the total
charge corresponding to their linear
superposition~(\ref{eq:double-kink}) does not necessarily equal $2Q$.
Therefore, instead of Eq.~(\ref{eq:double-kink}), we need to consider
a corrected configuration
\begin{eqnarray}
  {\bf u}(x)&=&{\bf u}^0(Q-\delta Q; x-x_1)+{\bf u}^0(Q-\delta
  Q;x_2-x)-{\bf m}\nonumber\\
  &=&{\bf u}^0(x-x_1)+{\bf u}^0(x_2-x)-{\bf m}+\delta {\bf
  u}(x),\label{eq:double-kink-corrected}
\end{eqnarray}
where the additional exponentially-small (of the order of the tail
overlap $\delta Q$) deformation
$$
\delta {\bf u}=- \delta Q\left[{\partial {\bf u}^0(Q;
      x-x_1)\over \partial Q}+{\partial{\bf u}^0(Q
  ;x_2-x)\over \partial Q}\right]
$$
serves to adjust the value of charge constraint, so that, {\em
  e.g.},
$$
\delta Q_L=\int_{-\infty}^{x_0}({\bf u}_2 -{\bf m}+\delta {\bf
  u})\, {\partial \rho({\bf u}_1)\over \partial {\bf u}_1} dx=0,
$$
and a similar condition for the region $x>x_0$ where the field
${\bf u}_2$ is far from equilibrium value (all notations as in
Sec.~\ref{sec:interaction}).  In the presence of the charge constraint
the Euler-Lagrange equations for a single kink must be written with a
chemical potential $\mu$, 
$$
 {}- {\bf u}_0''+ {\partial \over \partial {\bf u}}\left.\left[V({\bf
    u})+\mu\rho({\bf u})\right]\right|_{{\bf u}={\bf u}_0}=0,
$$
and the combination in the square brackets in the integrand of
Eq.~(\ref{eq:double-kink-energ}) no longer disappears.  Instead, it
changes the energy by an amount proportional to the total charge
increment $\delta Q_L$ in the region $x<x_0$, and a similar term for
$x>x_0$.  These charge increments vanish for the corrected
configuration~(\ref{eq:double-kink-corrected}), and in the linear order
we are again left with the same universal
expression~(\ref{eq:interaction-energy-simple}).  As before, it was
important that the correct configuration deviates very little from the
simple-minded superposition~(\ref{eq:double-kink}), including the tail
regions, where the correction $\delta{\bf u}$ can be safely ignored as
an exponentially small quantity of higher order.

Such linear analysis is
equivalent to finding the {\em instantaneous\/}
acceleration\cite{Manton-77,Jersak-77} of a defect surrounded by a
surface by calculating the total flux of the energy-momentum tensor
into the enclosed volume due to all other defects located outside the
surface.  The corrections to Eq.~(\ref{eq:interaction-energy-simple})
are easy to find in equilibrium, and they indeed turn out to be
exponentially smaller, if the locally-stable configuration of several
defects exists (in some cases such configurations can be stabilized by
the boundary conditions).  Often, however, because of the attraction
between individual solitons, there are no locally stable equilibrium
configurations minimizing the free energy.  In such cases, instead of
analyzing the forces in static configurations, the interaction can be
found more accurately by studying the full dynamics of the
system\cite{Abanov-Pokrovsky-1998}.  In the present work, however, we
are mostly interested in the {\em sign\/} of the interaction between
defects, and the accuracy of Eq.~(\ref{eq:interaction-energy-simple})
is sufficient.

\section{Untwisting instability of charged defects}
\label{app:untwisting}

Here we extend the local stability analysis of
Sec.~\ref{sec:cont-symmetry} to systems with conserved charge and
long-range interactions.  Now, instead of
Eq.~(\ref{eq:spin-rotat-part}), the relevant part of the free energy
and the corresponding constraint can be written as
\begin{eqnarray}
  {\cal F} &=& \!\!\int\!\chi({S}^2,x)\left({\nabla {\bf
        s}}\right)^2\!\!
  +V({S}^2,x)\,d^Dx\nonumber \\ 
  & &   +{1\over 2}\int\! \delta\rho(S^2,x)\,
        K(x,x')\,\delta\rho({S'}^2,x')\,d^Dx\,d^Dx',  
  \label{eq:one-component-fe}\\
  \!\! & &\int [\rho(S^2,x)-\bar\rho]\,d^Dx=0 ,
  \label{eq:one-component-constraint}
\end{eqnarray}
where the explicit coordinate dependence of the local part of the
potential energy $V$ and the charge density increment
$\delta\rho(S^2,x)\equiv \rho(S^2,x)-\bar\rho$ 
account for the presence of all other components $u_i^{(0)}(x)$, $2\le
i \le N$ of the order parameter.  The
expansion~(\ref{eq:gradient-energy-decomposition}) remains valid even
in the present case, and we can always select the ground-state
configuration of the functional~(\ref{eq:one-component-fe}) to have
only one component, ${\bf s}=(s_0(x),0)$.  As before, our task is to
prove that this configuration is locally unstable to ``untwisting'',
as long as the function $s_0(x)$ has a node.  The problem with the
charge constraint~(\ref{eq:one-component-constraint}) is slightly more
difficult, since the na\"\i vely perturbed configuration ${\bf
  s}=(s_0,s_1)$ generally has a different value of charge.  To correct
this, we consider a perturbed solution of the form
\[
{\bf s}=\{s_0\sqrt{1-\epsilon_1},\,\epsilon_2 w\},\quad
S^2=s_0^2+\epsilon_2^2 w^2-\epsilon_1\, s_0^2,
\]
where $\epsilon_1$ must be chosen to preserve the average charge
density, {\em i.e.},
\begin{equation}
  \label{eq:quadratic-order-constraint}
  \epsilon_1 = - \epsilon_2^2 
\left[{ \int \rho'\, w^2 d^Dx}\right]\,\left[{\int \rho' s_0^2 dx}
\right]^{-1} 
\end{equation}
where we assume that the denominator does not vanish identically, and
the derivative $\rho'\equiv \partial \rho(S^2,x)/\partial (S^2)$.

To quadratic order in $\epsilon_2$, the increment of the energy
functional~(\ref{eq:one-component-fe}) is just
\begin{eqnarray}
  \delta {\cal F}&=&\int dx\,\biggl\{
  \chi_0(x)\left[\epsilon_2^2(\nabla w)^2-\epsilon_1(\nabla
  s_0)^2\right]\nonumber\\
  & &\hskip-0.2in +\left[\epsilon_2^2\,w^2-\epsilon_1\,
  s_0^2\right]\,\left[\rho_0'\varphi_0(x)+V_0'+(\nabla
  s_0)^2\,\chi_0'\right]\biggr\}, 
  \label{eq:increment-ok}
\end{eqnarray}
where all functions with subscript $0$ are evaluated with the
non-perturbed configurations $s_0$, the prime denotes the derivatives
over $S^2$ as in Eq.~(\ref{eq:quadratic-order-constraint}), and the
scalar potential
$$
\varphi_0(x)\equiv\varphi([s_0^2],x)\equiv \int K(x,x')\,
\delta\rho(s_0^2(x'),x')\,d^Dx'. 
$$
Eq.~(\ref{eq:increment-ok}) can be simplified with the help of the
relation~(\ref{eq:quadratic-order-constraint}) and the
 Euler-Lagrange equation for the non-perturbed
solution $s_0$,
\begin{equation}
  \label{eq:self-adjoint-eq}
  -\nabla(\chi_0 \nabla
  s_0)+G([s_0^2],x)\,s_0=0,
\end{equation}
where the self-consistent potential function
$$
G([v_0^2],x)\equiv 
(\varphi_0(x)+\mu)\,\rho_0'+V_0'+(\nabla s_0)^2\,\chi_0'
$$
contains the Lagrange multiplier $\mu$.  We obtain, with the same
accuracy, 
\begin{eqnarray}
  \label{eq:fulldeltaF}
  \delta {\cal F}
  &=&\epsilon_2^2  \int d^Dx  \left[ \chi_0(\nabla w)^2 + 
    G([s_0^2],x) \,w^2\right]  . 
\end{eqnarray}

Let us return to the Euler-Lagrange
equation~(\ref{eq:self-adjoint-eq}).  As it stands, it is a non-linear
integro-differential equation for $s_0$.  However one can formally
look at this expression as an action of the linear self-adjoint
operator $\hat L = -\nabla(\chi_0\nabla)+G_0(x)$ (with fixed functions
$\chi_0(x)$ and $G_0(x)\equiv G([s_0^2],x)$) on the function $s_0$.
{}From this point of view $s_0$ is an eigenfunction of this operator,
$\hat L s_0 = \Lambda_0 s_0$ with zero eigenvalue $\Lambda_0 = 0$.
The same operator serves as the kernel of the energy
increment~(\ref{eq:fulldeltaF}), and so, expanding $w=\sum
A_l\,s_l(x)$ over the orthogonal eigenfunctions of this operators, we
obtain
$$
\delta {\cal F}=\epsilon_2^2 \sum \Lambda_l A_l^2 \int s_l^2\,d^Dx. 
$$
By assumption, $s_0$ has a node, and so there
must\cite{CourantHilbert} exist an eigenfunction $s_{-1}$ corresponding
to a negative eigenvalue $\Lambda_{-1}<0$.  Therefore, 
taking $w=s_{-1}$, we can decrease the free energy, 
$$
\Delta{\cal F}=\Lambda_{-1}\, \epsilon_2^2 
\int v_{-1}^2\,d^Dx   < 0, 
$$
which violates the original assumption.  Therefore, the spin
configurations with nodes are locally unstable to untwisting even in
systems with charge constraint and/or long-range interactions.

\section{No-node theorem for discrete systems}
\label{app:discrete}

It is also possible to prove a version of the no-node theorem for many
lattice models.  Consider the problem of finding a minimum of the
following expression
\begin{equation}
  \label{eq:discrete-energ}
  {\cal H}=\sum_{ij} \chi_{ij}\,({u}_i-{u}_j)^2+
  V({u}_1^2,\ldots,{u}_N^2),
\end{equation}
where the connections $\chi_{ij}\ge 0$ can be positive or zero, with
the only limitation that all points are linked, the variables ${u}_i$,
$i=1,\ldots,N$ are assumed to be scalars,\cite{endn:seven} and the
non-local potential $V({u}_1^2,\ldots,{u}_N^2)$ is a limited,
continuously differentiable function of all its arguments.  We are
going to prove that in the minimum of Eq.~(\ref{eq:discrete-energ})
all variables $u_i$ are non-zero and have the same sign, or all of
them vanish identically.

Let us suppose that the opposite statement is true, namely, that the
global minimum ${\cal H}^{(0)}$ is achieved on the set ${u}_i^{(0)}$,
some of which could be positive, negative or zeros, but at least one
non-zero value exists.  Without limiting generality, we can
suppose that this value is positive.  Let us now replace the original
set by the non-negative set $u_i^{(1)}=|u_i^{(0)}|$.  Clearly, because
of the obvious inequality (Cauchy)
$$
\left(a- b\right)^2\ge \left(| a|-| b|\right)^2, 
$$
this substitution cannot increase the energy.  This inequality
becomes strict if $a$ and $b$ have opposite signs, which implies that
the points with positive and negative values in the original
configuration must be separated by zeros, or our assumption was wrong.
Therefore, some of the values in the modified set $u_i^{(1)}$ are
expected to be zeros.  By assumption, there are no disconnected
points, and at least one point $j$ with zero value
$u_j^{(1)}=u_j^{(0)}=0$ must be connected to a point $i$ with
$u_i^{(1)}>0$.  If we replace the zero by a sufficiently small value
$u_j^{(2)}=\epsilon>0$, the increment of the
energy~(\ref{eq:discrete-energ}) will be negative, 
\begin{eqnarray}
  \label{eq:inequality}
  \delta{\cal H}_j&=&\sum_{i}\left\{\chi_{ij}\left(\epsilon^2
    -2\epsilon\,u^{(1)}_i\right)\right\}
  +\epsilon^2\,{\partial V({u}_1^2,\ldots)
    \over \partial {u_j^2}}\biggl|_{u_l=u_l^{(1)}}
  \nonumber\\
  &=&-2\epsilon\sum_i \chi_{ij} \,u^{(1)}_i+{\cal O}(\epsilon^2)<0. 
\end{eqnarray}
The procedure can be repeated for all points with zero value.
Therefore, the original assumption was wrong, and in the global minimum
all values $u_i$ must have the same sign (although they can be {\em
  exponentially\/} small).

Because the increment~(\ref{eq:inequality}) of the energy is {\em
  linear\/} in $\epsilon$, the proven statement can be easily extended
to accommodate an arbitrary dependence of the connections
$\chi_{ij}(u_1^2,\ldots,u_N^2)$ on the variables, as well as an
arbitrary number of non-local constraints of the form
$A(u_1^2,\ldots,u_N^2)=0$.


\begin{thebibliography}{10}
\bibitem{endn:two} {Although the long-distance interaction favors
    uniformly-charged configurations, one can show that periodic
    configurations necessarily replace the phase separation as long as
    this interaction is sufficiently weak.}
\bibitem{endn:three}
{A generalization to systems with charge constraint is considered in
  Appendix~\protect\ref{app:charged-interact}; also see related analysis in
  Sec.~\protect\ref{sec:antiphase}.}
\bibitem{endn:five} It was substantial that all physical quantities
  are continuously differentiable functions of $s^2$, while most
  generally the time-reversal symmetry only requires the dependence on
  the magnitude $|s|$. Such requirement is never a problem if $s$ is a
  component of a magnetic order parameter. In general, continuity
  implies that the proven statement can be violated only if the lowest
  eigenvalue of Eq.~(\protect\ref{eq:self-adjoint-ev}) diverges in a
  singular limit, $\Lambda_{-1}\to -\infty$.
\bibitem{endn:six} Invariants quadratic in $S_i$ and linear in
  derivatives also exist for seven point groups $C_{6v}$, $C_{4v}$,
  $C_{3v}$, $C_{2v}$, $D_{2d}$, $S_4$, and $C_s$ which do not contain
  inversion. Three remaining groups without inversion, $T_d$,
  $D_{3h}$, and $C_{3h}$, admit only quartic in $S_i$ and linear in
  derivatives invariants.
\bibitem{endn:seven}
The generalization to a multi-component order parameter is straightforward.
\end{thebibliography}
\end{document}